\documentclass{aastex62}
\usepackage[utf8x]{inputenc}
\graphicspath{{./}{figures/}}

\received{\today}
\revised{XXXXXX}
\accepted{XXXXX}
\submitjournal{ApJ}

\shorttitle{Born in a pair (?): Pisces~II and Pegasus~III}
\shortauthors{Garofalo et al.}

\begin{document}

\title{Born in a pair (?): Pisces~II and Pegasus~III\footnote{∗ Based on data collected with the Large Binocular Cameras at the Large
Binocular Telescope, PI: G. Clementini.}}

\correspondingauthor{Alessia Garofalo}
\email{alessia.garofalo@inaf.it}

\author[0000-0009-9999-9999]{Alessia Garofalo}
\affil{INAF-Osservatorio di Astrofisica e Scienza dello Spazio di Bologna, via Gobetti 93/3, 40129, Bologna, Italy}
\affil{Dipartimento di Fisica e Astronomia-Universit\'a di Bologna, via Gobetti 93/2, 40129 Bologna, Italy}

\author{Maria Tantalo}
\affiliation{Dipartimento di Fisica e Astronomia, Universit\'a degli Studi di Roma Tor Vergata, Via della Ricerca Scientifica 1, I-00133, Roma,
Italy}
\affil{INAF-Osservatorio Astronomico di Roma, Via Frascati 33, I-00078, Monte Porzio Catone,
Italy}

\author{Felice Cusano}
\affil{INAF-Osservatorio di Astrofisica e Scienza dello Spazio di Bologna, via Gobetti 93/3, 40129, Bologna, Italy}

\author{Gisella Clementini}
\affil{INAF-Osservatorio di Astrofisica e Scienza dello Spazio di Bologna, via Gobetti 93/3, 40129, Bologna, Italy}

\author{Francesco  Calura}
\affil{INAF-Osservatorio di Astrofisica e Scienza dello Spazio di Bologna, via Gobetti 93/3, 40129, Bologna, Italy}

\author{Tatiana Muraveva}
\affil{INAF-Osservatorio di Astrofisica e Scienza dello Spazio di Bologna, via Gobetti 93/3, 40129, Bologna, Italy}

\author{Diego Paris}
\affil{INAF—Osservatorio Astronomico di Roma, via Frascati 33, I-00078 Monte Porzio Catone, Italy}
\author{Roberto Speziali}
\affil{INAF—Osservatorio Astronomico di Roma, via Frascati 33, I-00078 Monte Porzio Catone, Italy}




\begin{abstract}

We have used $B,V$ time series photometry collected with the Large Binocular Telescope to undertake the first study of variable stars in the Milky Way ultra-faint dwarf (UFD) satellites, Pisces~II and Pegasus~III. In Pisces~II we have identified a RRab star, one confirmed and a candidate  SX Phoenicis star and, a variable with uncertain classification. In Pegasus~III we  confirmed the variability of two sources: an RRab star and a variable with uncertain classification, similar to the  case found in Pisces~II. Using  the  intensity-averaged apparent  magnitude  of  the  bona-fide  RRab  star in each galaxy we estimate    distance moduli  of  $(m-M)_{0}$= 21.22 $\pm$ 0.14 mag (d= 175 $\pm$ 11 kpc)  and  21.21 $\pm$ 0.23  mag (d=174 $\pm$ 18 kpc) for Pisces~II and Pegasus~III, respectively. 
Tests performed to disentangle the actual nature of variables with an uncertain classification led us to conclude that they most likely are bright, long  period and very metal poor RRab  members of their respective hosts. 
This may indicate that Pisces~II and Pegasus~III contain a dominant old stellar population (t$>$12 Gyr) with metallicity $<[Fe/H]>-$1.8 dex along with, possibly, a minor, more metal-poor component, as supported by the $V$, $B-V$ color-magnitude diagrams of the two UFDs and their  spectroscopically confirmed  members.  
The metallicity spread that we derived from our data sample is  $\gtrsim$0.4 dex in both systems.
Lastly, we built isodensity contour maps which do not reveal any irregular shape, thus making the existence of a physical connection between these UFDs unlikely.


\end{abstract}

\section{Introduction}\label{sec:intro}
Extensively under investigation in the last couple of decades, 
the mechanisms leading to the Milky Way (MW) formation and evolution 
and the role played by the MW dwarf spheroidal (dSph) satellites 
in assembling the Galaxy 
we observe today is still a very hot topic of modern astronomy. 
Under the current $\Lambda$CDM paradigm,
these old satellites could be survivors of the accretion events that led to the formation of the stellar halos of the large galaxies they orbit around (\citealt{bul05,sal07,sti17}).
Therefore, the MW  dSphs are privileged laboratories, contributing to our understanding of the Universe on both local and cosmological scales.
The MW dSphs are close enough to be resolved in stars, thus allowing to study the building up of the MW halo by exploiting information arising from the stars by which they are composed. In the last twenty years, observations carried out by wide-field surveys like the Sloan Digital Sky Survey (SDSS; \citealt{yor00}) have increased extraordinarily the numbers of companions detected around the MW. 
Since 2005 the SDSS has revealed that the MW is surrounded by  a new class of satellites, the ultra-faint dwarf (UFD) galaxies. UFDs  have such low stellar densities to make them very hard to detect unless specific techniques are applied to very large datasets as those provided by the SDSS. This survey in first place, then followed by other imaging surveys such as the Dark Energy Survey (DES; \citealt{des05}), Pan-STARRS1 3$\pi$ (\citealt{cha16}), ATLAS (\citealt{sha15}) and  strategic programs like the Subaru Hyper Suprime-Cam Survey (HSC; \citealt{aih18}), have increased the number of new MW satellites (UFDs and stellar clusters) up to 47 (\citealt{drl15,lae15,tor16,hom19} and references therein).
UFDs (see \citealt{sim19} for a 
state-of-the-art of the studies   on UFDs)
are characterised by low surface brightness $\mu_{V}\geq 28$ mag arcsec$^{-2}$ and luminosities $-8\leq M_V\leq -$1.5 \citep{mar08}, hence resulting fainter than all previously known dSphs (hereafter, classical dSphs) and the bulk of Galactic  globular clusters (GCs). On the absolute visual magnitude versus half-light radius ($M_V$-$r_h$) plane, they populate the extension to lower luminosities of the classical dSphs (see, e.g., figure~8 in \citealt{bel07} and figure~3 in \citealt{cle10}). These systems are believed to be strongly dark matter dominated, as their mass-to-light ratios and velocity dispersions are much higher than for the classical MW satellites (\citealt{str08}).
 Many UFDs present an irregular shape often interpreted as an evidence for tidal interaction with the MW \citep{bel07,mun10}, however their distorted morphologies seem to be more due to lack of deep enough observations rather than the signature of MW tidal stripping \citep[][and references therein]{sim19}.
It has been shown that the metal content in the UFDs is lower than in the classical dSphs and very similar to the chemical composition of  extremely metal-poor Galactic halo stars and the most metal-poor Galactic GCs  (e.g. \citealt{kir08}) making them likely the direct descendants of the first generation of galaxies in the Universe \citep{bov09,sal09}.
 All UFDs, but Leo~T \citep{cle12}, have no signature of ongoing star formation activity and display significant spreads in metallicity spanning [Fe/H] values down to $-$4 dex (see \citealt{tol09} and references therein).
They typically host metal poor ([Fe/H] $\leq-$ 2 dex) populations older than 10 Gyr as revealed by accurate spectroscopic studies of their stellar content (see e.g.  \citealt{seg07}, \citealt{kir08}, \citealt{nor10}, \citealt{sim11}, \citealt{kop15}).
Consistently with being populated by an old  stellar population and, with the only exceptions  of Carina~III \citep{tor18} and Willman~I \citep{sie08}, the UFDs which were studied for variability so far have been found to contain at least one RR Lyrae star. RR Lyrae variables (RRLs) are primary distance indicators and tracers of the oldest stellar population in the systems where they reside. 
RRLs oscillating in the fundamental  (RRab) and first overtone
(RRc) pulsation modes   
occupy two different and well-defined regions in the period-amplitude diagram (Bailey diagram;  \citealt{bai02,bai13,bai19}). According to the mean period of RRab stars ($\langle P_{ab}\rangle$) and the  fraction of
RRc stars, the RRLs in the MW field and GCs divide into Oosterhoff type I (Oo~I) and Oosterhoff type II (Oo~II) groups; this phenomenon is  known in the literature as Oosterhoff dichotomy \citep{oos39}. The Oo~I systems are more metal-rich ([Fe/H]  $\sim-$1.5 dex), their RRLs have shorter   periods ($\langle P_{ab}\rangle\sim$0.55 d) and a fraction of  RRc stars around 17$\%$. The Oo~II systems are mostly metal-poor ([Fe/H]$\sim-$2.0 dex) with longer RRab   periods  ($\langle P_{ab}\rangle\sim$0.65 d), bluer horizontal branches (HBs) than the Oo~I systems and a higher fraction of RRc stars ($\sim$ 37$\%$). 
 The properties of the RRLs and the Oosterhoff types observed in the MW halo (dSphs, GCs and field) can put constrains on the sub-structures which  contributed in the past to the formation of the MW.
The photometric studies carried out to define the pulsation properties of the RRLs in the MW satellites have revealed that the classical dSphs do not generally  conform to  the Oosterhoff dichotomy and, with the exception of Sagittarius and Ursa Minor, are all classified as Oosterhoff-Intermediate (\citealt{mat98}). Conversely, the UFDs for which a study of the variable stars has been carried out and RRLs have been  identified (21 UFDs so far),  although  containing only very few such variables  (hence, a robust Oosterhoff classification is not possible),  mainly tend to host RRLs with Oo~II properties, as observed for  the variables in the MW GCs and halo (\citealt{cle10}). 
\par
In this paper we present a study of the stellar population and variable star content of the  Pisces~II (Psc~II)  and Pegasus~III (Peg~III) UFDs, based on $B,V$ timeseries photometry obtained with 
the Large  Binocular Telescope (LBT). 
The  wide-field of view of the two cameras of the  LBT  allowed us to image an area corresponding to about 9 half-light radii in each individual pointing
of  Psc~II  and of about 12 half-light radii for  Peg~III.
 We also   investigate the possible physical connection between the two systems from the comparison of their properties (distance, CMDs and variable star populations). 
\par
The paper is organised as follows. In 
Sect.~\ref{sec:psc-peg} we 
present an overview of our targets. 
In Sect.~\ref{sec:psc_obs} we describe the   observations and the data reduction. From Sect.~\ref{sec:psc2_var} to~\ref{sec:v4_psc2} we discuss results from the identification and characterization of the variable stars, the Oosterhoff classification and measure the distance to Psc~II and Peg~III using the RRLs.
In Sect.~\ref{sec:psc2_cmd} we compare the CMDs of Psc~II and Peg~III with stellar evolution models. In Sec.~\ref{sec:metalR_psc2} we discuss the capability to retain and loss metals in low-mass galaxies such as our targets and, in Sect.~\ref{sec:iso_psc2} we present isodensity contour maps. 
Finally, Sect.~\ref{sec:conc_psc2} summarizes our results.
\section{Pisces~II and Pegasus~III}\label{sec:psc-peg}
The Psc~II UFD [R.A.(J2000) = 22h58m31s, DEC.(J2000) = 5$^\circ$57$^\prime$9$^{\prime\prime}$; l = 79.21$^\circ$, b = $-$47.11$^\circ$] was discovered in the southern Galactic portion of the SDSS Segue survey (\citealt{yan09}) by \citet{bel10}, and later studied in more detail by \citet{san12} and \citet{kir15}. Using deep wide-field photometry obtained with Megacam at the Magellan Clay telescope, \citet{san12} estimated the structural and photometric parameters of Psc~II.
From the mean magnitude of the HB they derived the distance modulus of 21.31 $\pm$ 0.17 mag,  corresponding to a heliocentric distance of 183 $\pm$ 15 kpc. \citet{san12} also estimated a half-light radius of 1.1 $\pm$ 0.2 arcmin, which delimits the region containing the bulk of the Psc~II stars and,  at the  estimated heliocentric distance,  corresponds to a linear extension of 58 $\pm$ 10 pc and an absolute magnitude M$_V$ = $-$4.1 $\pm$ 0.4 mag  for Psc~II. 
From the comparison of the galaxy color-magnitude-diagram (CMD) with theoretical isochrones by  \citet{gir04} and  \citet{dot07,dot08}, 
\citet{san12} concluded that Psc~II hosts a dominant old ($>$ 10 Gyr) and metal-poor ([Fe/H]$\sim$ $-$2 dex) stellar population. \citet{kir15} carried out a spectroscopic study of Psc~II using the Keck DEIMOS spectrograph. Their radial velocity measurements confirm the membership of seven candidate RGB stars, 
leading to a systemic velocity $<v_{\bigodot}>$ = $-$226.5 $\pm$ 2.7 kms$^{-1}$. They also measured spectroscopic metallicities for four of the seven confirmed members (see Table~\ref{tab:psc_mem},  upper portion), finding an average value of $\langle \rm{[Fe/H]} \rangle $ = $-$2.45 $\pm$ 0.07 dex and a metallicity
dispersion of $\sigma$[Fe/H]=0.48$\pm^{0.70}_{0.29}$.
Finally, \citet{kir15} estimated the following values for the velocity dispersion $\sigma$ = 5.4$^{+3.6}_{-2.4}$ kms$^{-1}$, dynamical mass log(M$_{1/2}$/M$_{\bigodot}$) = 6.2$^{+0.3}_{-0.2}$ and  mass-to light ratio M/L$_V$ = 370$^{+310}_{240}$ M$_{\bigodot}$/L$_{\bigodot}$, hence confirming that Psc~II is a dark matter dominated system.\\
\begin{table}[ht]
\caption{Members of Psc~II and Peg~III identified by literature spectroscopic studies.}
\centering
\scriptsize	
\label{tab:psc_mem}
\begin{tabular}{l c c c c}
\hline
\hline
\noalign{\smallskip}
    ID &RA (J2000)& DEC (J2000)& $v_{\bigodot}$& [Fe/H]\\
\hline   
Pisces~II \citep{kir15}\\
\hline 
M1-9004 &344.57314&5.92161&$-$224.9$\pm$1.6&$-$2.38$\pm$0.13\\
M2-9833 & 344.58857& 5.95573& $-$226.9$\pm$3.2 &...\\
M3-10694 & 344.60458 &5.95572& $-$232.0$\pm$1.6&$-$2.70$\pm$0.11\\
M4-12924 &344.64224&5.96213&$-$221.6$\pm$2.9&$-$2.10$\pm$0.18\\
M5-13387 &344.64976& 5.95625&$-$215.8$\pm$7.6&...\\
M6-13560 &344.65311&5.96841&$-$232.6$\pm$5.3&$-$2.15$\pm$0.28\\
M7-14179 &344.66411&5.93354&$-$224.8$\pm$9.6&...\\
\hline   
Pegasus~III \citep{kim16}\\
\hline 
M1 &336.09960&5.4118042&$-$226.16$\pm$5.04&$-$2.24$\pm$0.15\\
M2 & 336.11029& 5.4159619& $-$229.45$\pm$5.29 &$-$2.55$\pm$0.15\\
M3 & 336.11378 &5.4077794& $-$218.51$\pm$3.64&$-$2.55$\pm$0.15\\
M4 &336.09539&5.3958641&$-$234.68$\pm$3.84&...\\
M5 &336.10025& 5.424227&$-$218.26$\pm$3.56&$-$2.24$\pm$0.15\\
M6 &336.10207&5.3891321&$-$220.57$\pm$4.71&...\\
M7$^{*}$ &336.08666&5.39347&$-$193.35$\pm$22.92&...\\
M8 &336.07518&5.4196487&$-$208.45$\pm$6.66&...\\
 \noalign{\smallskip}
 \hline  
\hline    
\end{tabular}\\
\normalsize
\tablecomments{$^{*}$Uncertain membership.}
\end{table}

A few years later another new  MW UFD, Peg~III [ R.A.(J2000) = 22h24m48s, DEC.(J2000) = 5$^\circ$24$^\prime$18$^{\prime\prime}$; l = 69.85$^\circ$, b = $-$41.83$^\circ$] was  discovered from an analysis of the SDSS Data Release 10 (DR10; \citealt{ahn14}) and then confirmed with deeper Dark Energy Camera (DECam) follow-up observations by \citet{kim15}. 
The  physical properties derived for  Peg~III are very similar to those of other  previously known MW UFDs, and indeed, Peg~III perfectly locates in the UFD region of the $M_V$-$r_h$ diagram.
A further study of this galaxy was carried out by \citet{kim16}, who using deeper photometry and spectroscopy data from the Magellan/Baade Telescope Inamori-Magellan Areal Camera \& Spectrograph (IMACS) provided a better definition of the Peg III structural and photometric parameters.
By fitting the HB fiducial line of the Galactic GC  M15 ($\langle \rm{[Fe/H]} \rangle$ = $-$2.42 dex; \citealt{ber14}) to the blue horizontal branch stars, \citet{kim16} derived a distance modulus $(m-M)_{0}$ = 21.66 $\pm$ 0.12 mag corresponding to a heliocentric distance d$_{\bigodot}$ = 215 $\pm$ 12 kpc. Using spectroscopic data obtained with the Keck DEIMOS spectrograph, they also measured the radial velocity of a number of candidate members confirming the membership to Peg III of seven of them and obtaining a systemic velocity of $-$222.5 $\pm$ 2.6 kms$^{-1}$. A further candidate member has a radial velocity about 30 kms$^{-1}$ larger than the systemic velocity inferred from the other seven stars, hence its actual membership to Peg~III remains doubtful. The radial velocities of the 8 stars  are listed in Table~\ref{tab:psc_mem}(bottom panel).
The position of the spectroscopically confirmed member stars and the comparison of the CMD  with the Dartmouth isochrones \citep{dot08} led \citet{kim16} to conclude that Peg~III has a dominant old (13.5 Gyr) and metal-poor ($\langle {\rm [Fe/H]} \rangle$ = $-$2.5 dex) stellar population, in excellent agreement with  most of the MW UFD satellites.


The new values  of the half-light radius (r$_{h}$ = 0.85 $\pm$ 0.22 arcmin, which at the new heliocentric distance corresponds to r$_{h}$ = 53 $\pm$ 14 pc) and absolute magnitude ( M$_{V}$ = $-$3.4 $\pm$ 0.4 mag)  do not dramatically change the position of Peg III on the M$_{V}$-r$_{h}$ diagram, and keep the galaxy in the UFD region. 

\citet{kim16} measured the spectroscopic metal abundance of the four brightest stars among the confirmed members, obtaining metallicity estimations from $-$2.24 $\pm$ 0.15 dex to $-$2.55 $\pm$ 0.15 dex. These values are in good agreement with the metallicity inferred from the fit with the old and metal-poor Dartmouth isochrone. The velocity dispersion $\sigma_{v}$ = $-$5.4$^{+3.0}_{-2.5}$ km s$^{-1}$, the dynamical mass M$_{1/2}$ = 1.4$^{+3.0}_{-1.1}$ $\times10^{6}M_{\bigodot}$ and the mass-to-light ratio 
 M/L$_V$ = 1470$^{+5660}_{-1240}$ M$_{\bigodot}$/L$_{\bigodot}$ confirm that Peg~III is more  dark matter dominated   than Psc~II. 
In addition, \citet{kim16} provide a density contour map that shows the irregular shape of the Peg III structure, mostly in the outermost regions. This could be the sign of an on-going interaction. Finally,  \citet{kim15, kim16} noticed that 
Peg~III is spatially close to Psc~II. They 
 are separated by only 8.5$^\circ$ on the sky and have fairly similar distances. 
 The spatial separation of the two UFDs is $\sim$43$\pm$19 kpc (\citealt{kim16}). If Psc~II and Peg~III do indeed form a physical pair they could resemble the case of the Leo~IV and Leo~V,  
(\citealt{jon10}) for which an origin related to a single, tidally disrupted progenitor has been proposed.  However,  they could also resemble the couple Carina~II and Carina~III \citep{tor18}, which have a projected separation of only $18^\prime$ on sky, hence smaller than Leo~IV and Leo~V, but are not a pair of bound systems according to \citet{li18}. 
The parameters derived in previous studies of Psc~II (\citealt{kir15}, \citealt{san12}, \citealt{bel10}) and Peg~III \citep{kim16} are summarized in Table~\ref{tab:psc2_main}.

\begin{table}[ht]
\caption{Main parameters of Psc~II and Peg~III from literature studies.}
\centering
\label{tab:psc2_main}
\begin{tabular}{l c c c}
\hline
\hline
\noalign{\smallskip}
     Parameter &Pisces~II     & Pegasus~III & Ref.$^{a}$\\
\hline   
RA(J2000)& 22$^{h}$58$^{m}$31$^{s}$&22$^{h}$24$^{m}$24.48$^{s}$&1,4\\
DEC(J2000)& 5$^\circ$57$^\prime$9$^{\prime\prime}$&5$^\circ$24$^\prime$18$^{\prime\prime}$&1,4\\
d$_{\bigodot}$ (kpc)&183$\pm$15&215$\pm$12&1,3\\
M$_{V}$ (mag)&$-$4.1$\pm$0.4&$-$3.4$\pm$0.4&1,3\\
r$_{h}$ (arcmin)&1.1$\pm$0.2&0.85$\pm$0.22&1,3\\
r$_{h}$ (pc)&58$\pm$10&53$\pm$1.4&1,3\\
$\epsilon$&$<0.31$&0.38$^{+0.22}_{-0.38}$&1,3\\
$\theta$ (deg)&unconstrained&114$^{+19}_{-17}$&1,3\\
M$_{1/2}$ (10$^6$M$_{\bigodot}$)&6.2$^{+0.3}_{-0.2}$ $^{c}$&1.4$^{+3.0}_{-1.1}$&1,2\\
M$_{1/2}$/L$_{V}^{b}$ (M$_{\bigodot}$/L$_{\bigodot}$)&370$^{+310}_{-240}$&1470$^{+5660}_{-1240}$&1,2\\
$<v_{\bigodot}>$ (km s$^{-1}$)&$-$226.5$\pm$2.7&$-$222.9$\pm$2.6&1,2\\
$\sigma_{v}$ (km s$^{-1}$)&5.4$^{+3.6}_{-2.4}$&5.4$^{+3.0}_{-2.5}$&1,2\\
$v_{GSR}$ (km s$^{-1}$)&$-$79.9$\pm$2.7&$-$67.6$\pm$2.6&1,2\\
$<[Fe/H]>$ (dex)&$-$2.45$\pm$0.07&$-$2.55$\pm$0.15$^{d}$&1,2\\
&&$-$2.24$\pm$0.15$^{d}$&\\
\noalign{\smallskip}
 \noalign{\smallskip}
 \hline  
\hline    
\end{tabular}
\normalsize
\tablecomments{$^{a}$References: (1)\citet{kim16}; (2) \citet{kir15}; (3) \citet{san12}; (4) \citet{bel10} $^{b}$Mass-to-light ratio within the half-light radius.$^{c}$The value is referred to log(M$_{1/2}$/M$_{\bigodot}$). $^{d}$Individual values of spectroscopic metallicities derived by \citet{kim16},
as they do not provide the mean value.}
\end{table}

\section{LBT observations and PSF photometry of Psc~II and Peg~III}\label{sec:psc_obs}

$B$,$V$ time-series photometry of Psc~II was collected with the Large Binocular Cameras \citep[LBCs,][]{giallo2008} of the LBT, 
program (PI: G. Clementini) which targeted also the MW UFD Peg~III. Each LBC has a field of view (FoV) of 23$^{\prime}\times$23$^{\prime}$ and is equipped with four 2048 $\times$ 4608 pixel EEV CCDs resulting in a 6150 $\times$ 6650 pixel equivalent detector, with a pixel scale of 0.224$^{\prime\prime}$/pixel. The data consist of 26 $B$ and 26 $V$ images for Psc~II and of 28 $B$ and 28 $V$ images for Peg~III,  each with an exposure time of 180s, acquired over six nights from October to December, 2015. This exposure time allowed us to obtain a signal-to-noise ratio (S/N) $\geqslant$ 50 at $B\sim$22.5-23 mag,  which roughly corresponds to the minimum in the $B$ and $V$ light curves of RRLs, according to the Psc~II and Peg~III distance moduli.
Most of the observations were obtained with seeing $\leq$ 1.2 arcsec. This is well suited for our purposes. A log of the Psc~II and Peg~III observations is provided in Table~\ref{tab:psc2_log}.
All images were pre-reduced (bias-subtracted, flat-fielded and astrometrized) by the LBT team through an LBC dedicated pipeline, 
and then, on each of the four CCDs of the two LBCs, separately, we performed point spread function (PSF) photometry with the \texttt{DAOPHOT-ALLSTAR-ALLFRAME} packages (\citealt{ste87,ste94}). To transform the instrumental magnitudes to the Johnson standard system we used standard stars selected from the SDSS catalog and 
the transformation equations available at \url{https://www.sdss3.org/dr8/algorithms/sdssUBVRITransform.php#Lupton2005}. We used  calibration equations of the form: $B-b$= c$_B$+m$_B\times$($b-v$) and $V-v$=c$_V$+m$_V\times$($b-v$),  where $B$ and $V$ are the standard Johnson magnitudes of the SDSS stars, and $b$ and $v$ are the instrumental magnitudes in our photometric catalogs. The parameters of the calibration were derived by applying a 3$\sigma$ clipping rejection algorithm to fit the data. The final fit to calibrate the Psc~II catalog was obtained using a total of 772 stars and provided the following calibration equations: $B-b$=8.229$-$0.1409$\times$($b-v$) and 
$V-v$=8.139$-$0.0823$\times$($b-v$), with r.m.s values of 0.034 and 0.024 mag in $B$ and $V$, respectively. For Peg~III, the sample used for the final fit contains 669 stars and provided the following calibration equations:
$B-b$=8.226$-$0.1348$\times$($b-v$) and 
$V-v$=8.195$-$0.0728$\times$($b-v$),
with the r.m.s. values of 0.03 mag for the $B$ and 0.025 mag for the $V$ magnitudes, respectively. These calibration equations, derived independently for Psc~II and Peg~III, are fully consistent with each other.
Our final $B,V$ combined catalogs of sources observed in the LBT fields centered on Psc~II and Peg~III contain each more than 28,000 objects. The photometric errors for non variable sources in our catalogues  typically range from 0.0015 up to 0.01 mag for B $<$ 20.0 mag, from 0.01 up to 0.15 mag for 20.0 $<$ B $<$ 25 mag, then increase steeply for B $>$ 25 mag. In particular, the photometric errors at the HB level are of about 0.06-0.07 mag in $V$ and $B$ for both galaxies.\\ 
Unfortunately, an unexpected rotation of the LBT occurred during the observations of Peg~III. 
Therefore some portions of the FoV covered by our observations were deeper than others, causing the detection of fainter sources in those parts  of the sky. This produced a more populated CMD but also fictitious overdensities in the density maps, as we further discuss in Sect~\ref{sec:iso_psc2}.

\begin{table}[ht]
\caption{Log of observations}
\centering
\label{tab:psc2_log}
\begin{tabular}{l c c c c}
\hline
\hline
\noalign{\smallskip}
     Dates & Filter & N& Exposure Time& Seeing\\
     & & &(s)&(arcsec)\\
\hline   
\noalign{\smallskip}
     Pisces~II \\
 \noalign{\smallskip}
 \hline  
October 2, 2015 & B & 11 & 180 & 1.1-1.3\\
October 2, 2015 & V & 11 & 180 & 0.9-1.4\\
November 8, 2015& B & 2 & 180 & 1.3-2.2\\
November 8, 2015& V & 2 & 180 & 1.3-2.2\\
November 9, 2015& B & 6 & 180 & 0.7\\
November 9, 2015& V & 6 & 180 & 0.7\\
December 1, 2015& B & 4 & 180 & 1.1-1.3\\
December 1, 2015& V & 4 & 180 & 0.9-1.2\\
December 2, 2015& B & 3 & 180 & 1\\
December 2, 2015& V & 3 & 180 & 1.1-1.4\\  
 \hline    
 \noalign{\smallskip}
     Pegasus~III \\
\noalign{\smallskip}     
\hline 
October 2, 2015 & B & 5 & 180 & 1.1-1.3\\
October 2, 2015 & V & 5 & 180 & 0.9-1.4\\
November 7, 2015& B & 5 & 180 & 0.8-1\\
November 7, 2015& V & 5 & 180 & 0.8-1\\
November 8, 2015& B & 4 & 180 & 1.3-2.2\\
November 8, 2015& V & 4 & 180 & 1.3-2.2\\
November 9, 2015& B & 5 & 180 & 0.7\\
November 9, 2015& V & 5 & 180 & 0.7\\
December 1, 2015& B & 5 & 180 & 1.1-1.3\\
December 1, 2015& V & 5 & 180 & 0.9-1.2\\
December 2, 2015& B & 4 & 180 & 1\\
December 2, 2015& V & 4 & 180 & 1.1-1.4\\
\hline    
\end{tabular}\\
\normalsize
\end{table}

\section{Variable stars: identification and pulsation properties}\label{sec:psc2_var}
To identify variable stars in Psc~II and Peg~III we first considered stellar sources with  a high  value of the variability index computed by \texttt{DAOMASTER} (\citealt{ste94})  both in the $V$ and $B$ passbands.  Then we also checked all stars that in the CMDs of our  two targets have colors and magnitudes falling within the edges of the 
classical instability strip (IS).  
For Psc~II this procedure returned a list of $\sim$ 20 candidate variables with more than 22/24 data points on the light curves which were visually  inspected with the GRaphical Analyzer of TImes Series package (GRATIS, \citealt{clm00}). GRATIS uses the Lomb periodogram (\citealt{lom76}; \citealt{sca82}) to obtain a first guess of the star periodicity  and then the best fit of the folded light curve  with a truncated Fourier series to refine the star period.  
The final adopted periods were those that minimize the r.m.s scatter of the truncated Fourier series best fitting the data. Taking into account the  period, the shape of the light curve, the position on the CMD according to the intensity-averaged $B,V$ magnitudes, we confirmed the variability of three candidates and classified in type two of them: 1 RRab and 1 SX Phoenicis (SX Phe) star, whereas the classification of the third confirmed variable remains uncertain 
because two alternative periods are equally probable for the star. Correspondingly, this variable could  either be a metal poor RRab or an anomalous Cepheid (AC).
\begin{figure*}[ht]
\centering
\includegraphics[trim= 60 160 40 100 clip, width=14cm]{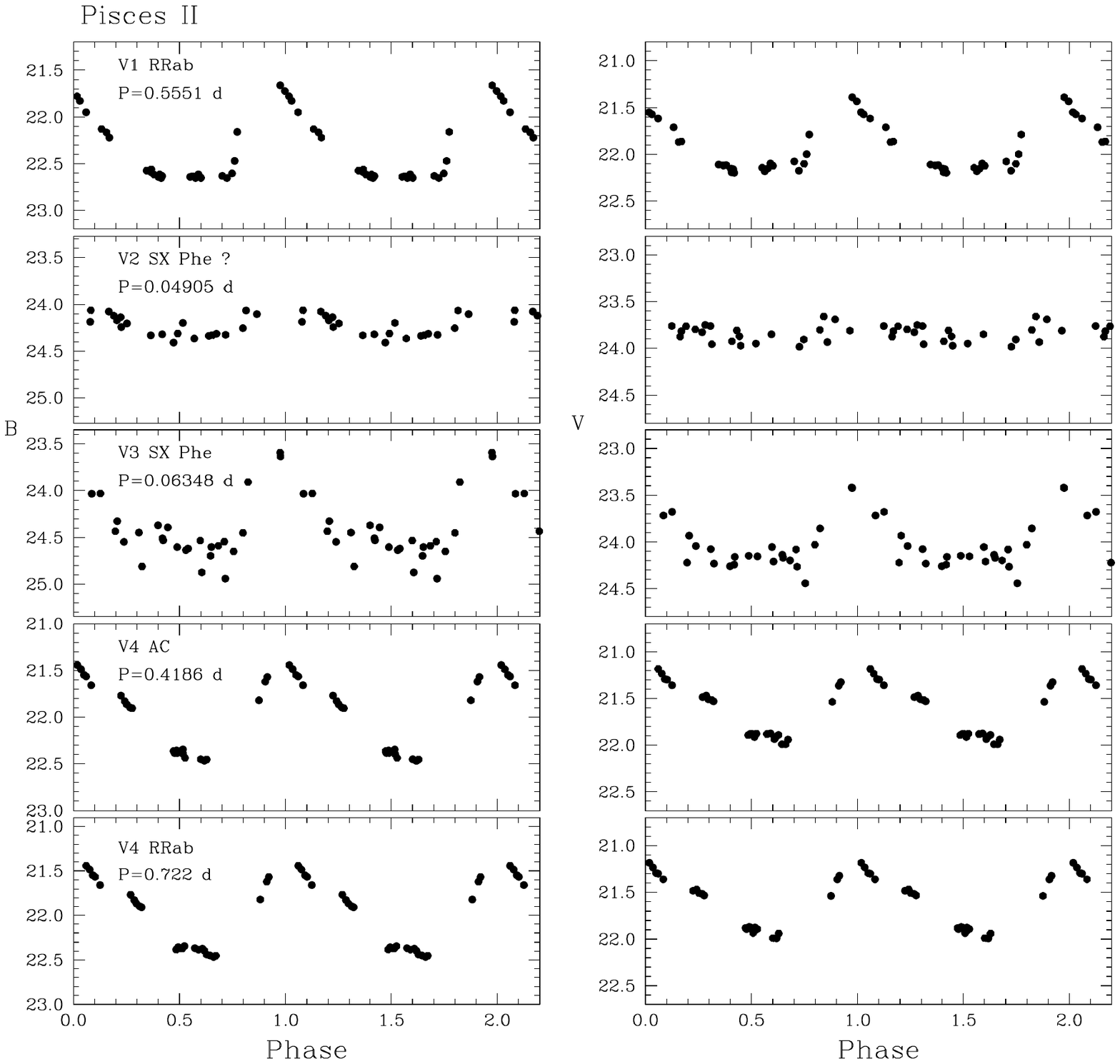}
\caption{$B$ (left panels) and $V$ (right panels) light curves of confirmed and candidate variable stars  identified in Psc~II. For V4 the data have been folded according to the two equally probable  periods (bottom two  panels). Typical internal errors of the single-epoch data range from  0.01 to 0.07 mag in $B$, and from 0.01 to 0.03 mag in $V$ for the bright variables (V1 and V4) and from 0.03 to 0.30 mag in $B$ and from 0.04 to 0.19 mag in $V$,  for the SX Phe stars (V2 and V3).}
\label{fig:lc_psc2}
\end{figure*}
\begin{figure}[hbp]
\centering
\includegraphics[trim= 60 150 40 70 clip, width=8cm]{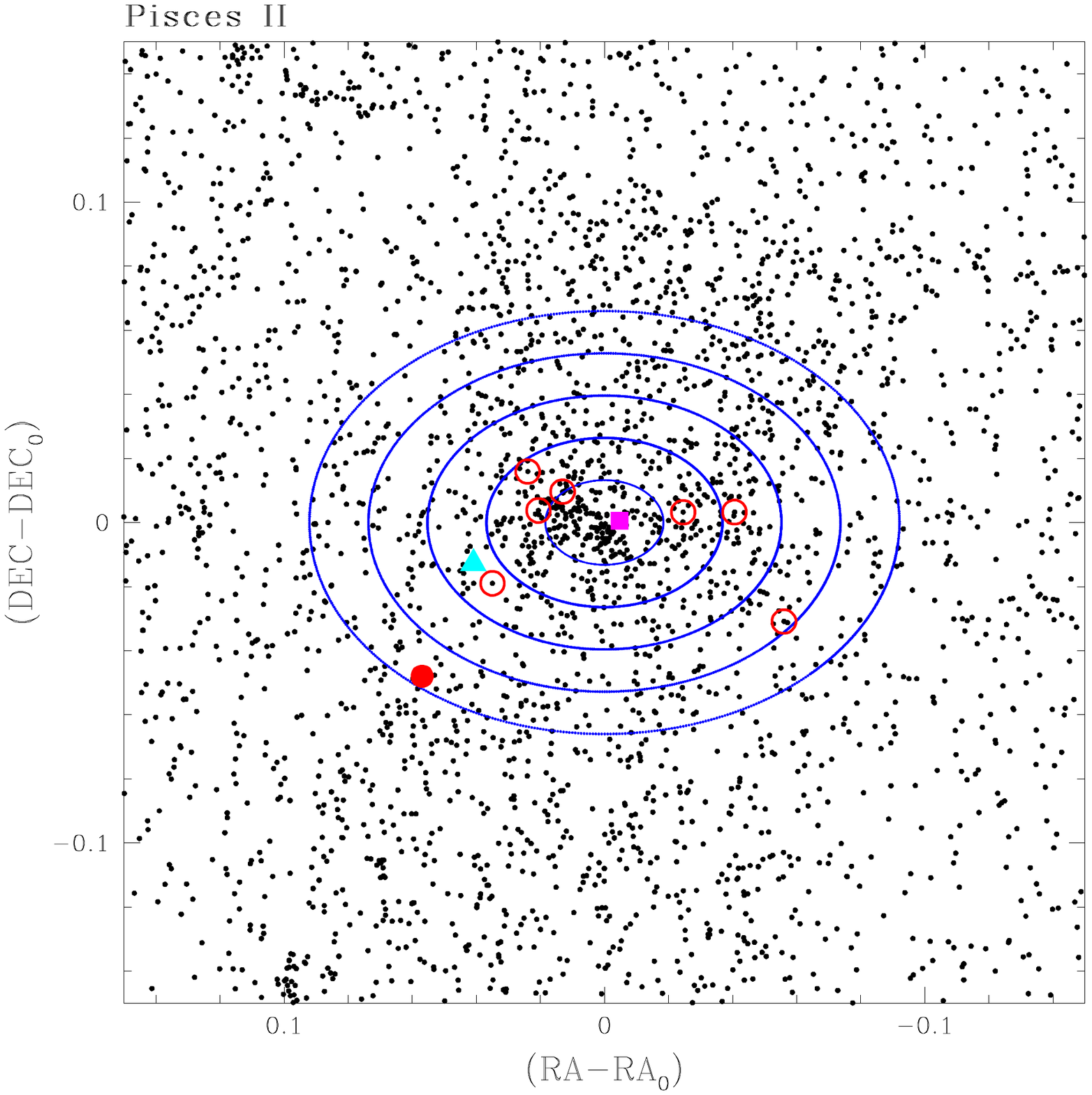}
\caption{Map of the sources measured in the LBC FoV of Psc~II with respect to the galaxy center coordinates. Only sources selected according to the cuts described in Sect.~\ref{sec:psc2_cmd} are displayed. Five blue ellipses are drawn showing from one (inner ellipse) to 5 times the galaxy r$_{h}$ (outer ellipse; that is where V4, the outermost variable of Psc~II is located) adopting the position angle, r$_{h}$, ellipticity, and center coordinates of \citet{san12}. A magenta filled square marks the fundamental mode RRL (V1), a cyan filled  triangle the confirmed SX Phe star (V3) and a red filled circle the RRab/AC star (V4). Red open circles show 7 members of Psc~II spectroscopically confirmed by \citet{kir15}.}
\label{fig:map_psc2}
\end{figure}
Since the available data did not allow us to firmly establish the correct periodicity, hence  classification in type, of this source, in this work we used both periodicities to fold the star light curves. A forth source, which is located inside the galaxy r$_h$, 
exhibits variability particularly in the $B$ passband. We tentatively classified this source as an SX Phe star, however, this classification remains doubtful because the $V$ light curve shows very little, if any, variability (see Fig.~\ref{fig:lc_psc2}). 
The $B$, $V$ time-series photometry of the variable stars identified in Psc~II is  provided in the upper portion of Table~\ref{tab:ts}. The properties of the variable stars are summarised in the upper portion of Table~\ref{tab:psc2_var}. 
Column 1 of Table~\ref{tab:psc2_var} provides the star identifier. This is an increasing number starting from the galaxy center, for which we have adopted the coordinates of the discovery paper (\citealt{bel10}). Columns 2 and 3 give the star coordinates.  Columns 4, 5 and 6 list respectively the classification in type, the period (P) and the Heliocentric Julian Day (HJD) of maximum light. Columns 7, 8, 9 and 10 provide the intensity-averaged mean $B$ and $V$ magnitudes and the corresponding amplitudes of the light variation. 
The $B$, $V$ light curves of the confirmed (3) and candidate (1) variable stars in Psc~II are shown in Fig.~\ref{fig:lc_psc2}.
The spatial distribution of all stars measured in our LBT observations of Psc~II with respect to galaxy center (RA$_0$, DEC$_0$) is shown in Fig.~\ref{fig:map_psc2}. In the figure we have marked in blue five ellipses representing from one to five times the r$_h$ of Psc~II, 1.1 $\pm$ 0.2 arcmin, as estimated by \citet{san12}. 
The bona-fide RRab star (V1; magenta filled square) is located within the Psc~II r$_{h}$, at a distance from the center of about 0.3$^{\prime}$. The SX Phe star (V3; cyan triangle) and the variable with  uncertain type (V4; red filled circle) both are outside the galaxy r$_{h}$, at distances of about 2.6$^{\prime}$ and 4.5$^{\prime}$ from the galaxy center. In particular, the SX Phe star is within 3 r$_{h}$ and the  variable with uncertain type within 5 r$_{h}$. The distance from the galaxy center and the position on the CMD (see Figure~\ref{fig:cmd_psc2_5r}) 
of bona-fide RRab and SX Phe stars suggest that they belong to Psc~II; conversely, 
the membership of V4 remains doubtful (see Section~\ref{sec:v4_psc2}). 
In Peg~III the same procedure as  applied for Psc~II    allowed us to  confirm the variability of two sources in our list of candidates: an  RRab star,  according to the period, the shape of the light curve, the position on the CMD and the $B$, $V$ mean  magnitudes; and a variable star with uncertain classification.  As for V4 in Psc~II GRATIS finds two equally possible periodicities for the Peg~III  variable with an uncertain classification. 
We used both periodicities to fold the star light curves. According to the two different periods, the corresponding amplitudes of the light variation and 
the $V$ mean magnitude being about 0.3 mag brighter than the HB level in Peg~III,
similarly to V4 in Psc~II, this variable could either be a very metal poor RRab or an AC. 
A few additional $B$ and $V$ measurements would be sufficient  to improve the light curve sampling and the period definition, thus allowing to definitively classify this variable. 
 The $B$ and $V$ time-series photometry and the properties of the two variable stars we have identified in Peg~III are provided in the lower portions of Table~\ref{tab:ts} and \ref{tab:psc2_var},  respectively. We have named them V1 and V2 according to their distance from the galaxy center for which we adopted the coordinates by \citet{kim16}. V1 corresponds to the RRab star and V2 is the variable of uncertain type.
Their $B$ and $V$ light curves are shown in Fig.~\ref{fig:lc_peg3}, where the data of V2 were folded using the periodicity as RRab in the middle panel and the periodicity as AC in the bottom panel.
Figure~\ref{fig:map_peg3} shows the spatial distribution of all the stars measured in our LBT observations with respect to the center of Peg~III. We have marked V1 with a red filled circle, V2 with a red open circle, spectroscopically confirmed members (according to \citealt{kim16}) with blue stars and an ambiguous member with an orange star. In the figure, five green ellipses indicate, in increasing order, 1, 2, 3, 5 and 10 times the half-light radius (r$_{h}$) of Peg~III, r$_{h}$ = 0.85 $\pm$ 0.22 arcmin as computed by \citet{kim16}. The two confirmed variable stars lie outside the galaxy’s r$_{h}$, at distances from the galaxy center of 2.5$\arcmin$ for the bona-fide RRab and 5.8$\arcmin$ for the uncertain type variable. The bona-fide RRab is within 4 r$_{h}$, and its position on the HB
(see 
right panel of Fig.~\ref{fig:cmd_psc2_5r}) 
supports its membership to Peg~III. On the contrary, V2 lies within 10 r$_{h}$ (the star was identified in the data set of CCD1) and is about 0.3 mag above the HB level (see 
right panel of  
Fig.~\ref{fig:cmd_psc2_5r}), thus casting  doubts on the actual membership to Peg~III of this variable. We discuss V2 more in detail in Section ~\ref{sec:v4_psc2}.
\begin{figure*}[ht]
\centering
\includegraphics[trim= 60 350 40 110  clip, width=14cm]{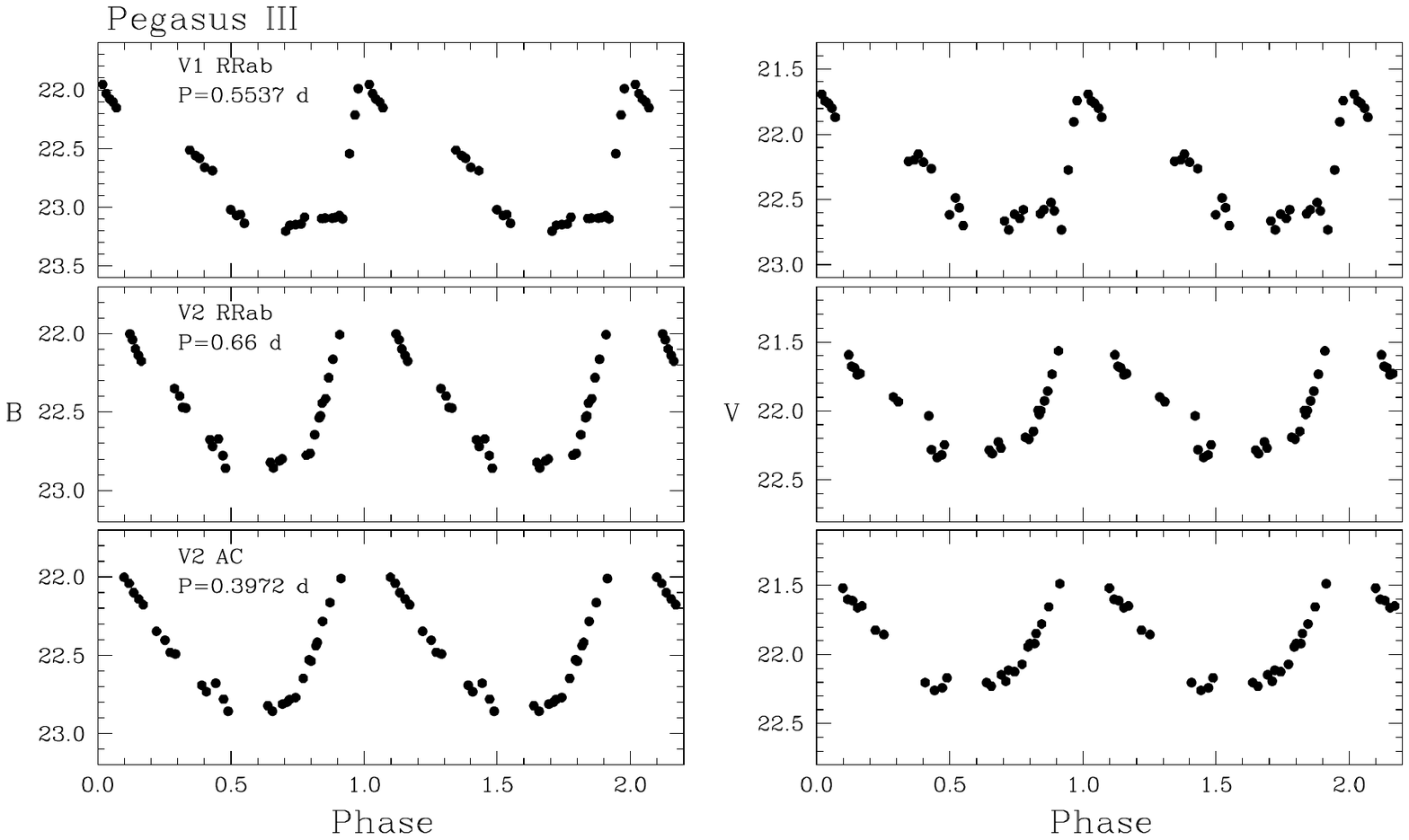}
\caption{$B$ (left panels) and $V$ (right panels) light curves of the two pulsating variable stars we have identified in 
Peg~III. As for V4 in Psc~II,
for V2 the data have been folded according to the two equally probable periods (center and bottom panels). Typical internal errors for the single-epoch data range from 0.01 to 0.08 mag in $B$, and from 0.01 to 0.06 mag in $V$.}
\label{fig:lc_peg3}
\end{figure*}
\begin{figure}[ht]
\centering
\includegraphics[trim= 0 150 0 0 clip, width=9.5cm]{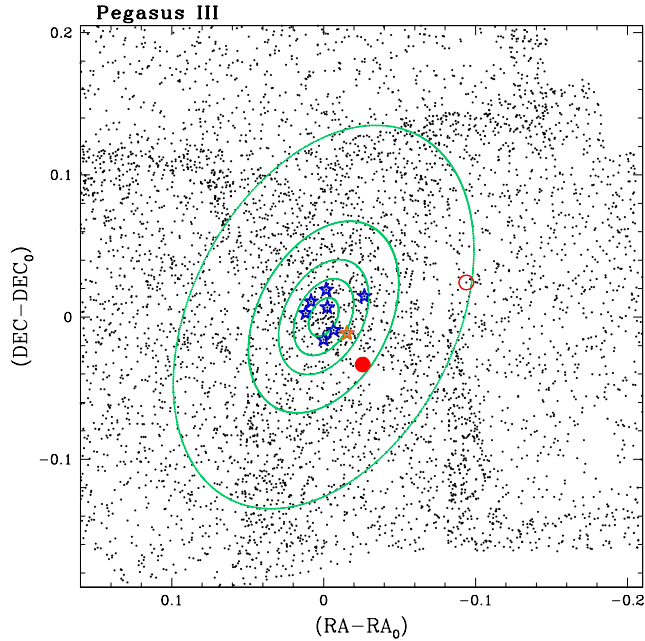}
\caption{
Map of the sources measured in the LBC FoV of Peg~III with respect to the galaxy center coordinates.  Only sources  selected  according  to  the  cuts  described  in  Sect.  8 are displayed.
We have marked the position of the two confirmed  variables  using a red filled circle for the RRab star (V1) and a red open circle for V2. We also show the spectroscopically confirmed members (blue stars) and an ambiguous member (orange star) according to \citet{kim16}. Five green ellipses were drawn convolving the Peg~III center coordinates, ellipticity, position angle, r$_{h}$, and its multiples, according to the values provided by \citet{kim16}. The five ellipses correspond to 1, 2, 3, 5 and 10 times the r$_{h}$ of Peg~III.}
\label{fig:map_peg3}
\end{figure}


\begin{table}[ht]
\caption{$B$ and $V$ timeseries photometry of the Psc~II and Peg~III variable stars. The table is published in its entirety in the electronic edition of the journal. A portion is shown here for guidance regarding its form and content.}
\centering
\label{tab:ts}
\begin{tabular}{l c c l c c}
\hline
\hline
\noalign{\smallskip}
     HJD & B & $\sigma_{B}$& HJD &$V$& $\sigma_{V}$\\
     ($-$2457000)& (mag)& (mag)&(-2457000)&(mag)&(mag)\\
\hline   
\noalign{\smallskip}
     Pisces~II - V1 \\
 \noalign{\smallskip}
 \hline  
335.60833 & 21.7788  &  0.040    & 335.60828 & 21.3891 &0.0297\\
335.62004 & 21.8276 &   0.036  &  335.61999 & 21.4334& 0.0275\\ 
335.63116 & 21.9492  &  0.030   &  335.63109&  21.5478& 0.0248\\
335.63837 & 22.1282 &   0.045 &   335.63831 & 21.5708&0.0270\\ 
335.65506 & 22.1653  &  0.065 &   335.655  &  21.6142 &0.0300\\
334.58568 & 22.2208 &   0.026 &   334.58562  & 21.7106  & 0.0180\\
334.59818 & 22.5726  &  0.039 &  297.7517  & 21.7876 & 0.0213\\
334.60584 &  22.5909 &   0.032    & 334.59814 &  21.8667 &0.0199\\
... & ... & ... &...&...&...\\
 \hline    

\end{tabular}\\
\normalsize
\end{table}

\begin{table*}[ht]
\caption{Properties of the variable stars identified in Psc~II and Peg~III}
\centering
\small
\label{tab:psc2_var}
\begin{tabular}{l c c c c c c c c c c}
\hline
\hline
\noalign{\smallskip}
     Name & RA     & Dec & Type & P      & Epoch (max)   & $\langle$B$\rangle$ & A$_{B}$ & $\langle$V$\rangle$ & A$_{V}$& [Fe/H]$^{*}$\\
          & (J2000)      & (J2000)  &      & (days) & HJD (2457000) &         (mag)       &  (mag)  &          (mag)      &  (mag) & (dex)\\
\hline   
\noalign{\smallskip}
     Pisces~II \\
 \noalign{\smallskip}
 \hline  
     V1   & 22:58:29.858 & +5:57:10.92 & RRab  & 0.5551  & 296.71  & 22.28 & 0.94 & 21.89 & 0.69&$>-1.6/-1.7$\\   
     V2   & 22:58:33.535 & +5:57:43.63 & SXPhe(?) & 0.04905     & ...      & ...    & ...   & ...    & ...\\    
     V3   & 22:58:40.807 & +5:56:23.16 & SXPhe & 0.06348 & 333.6245 & 24.245 & 1.119 & 23.935 & 0.775 \\    
     V4$_{\rm RRL}$   & 22:58:44.659 & +5:54:16.33 & RRab  & 0.722   & 335.565  & 21.98 & 1.18 & 21.63 & 0.93& $>-$2.3\\  
      V4$_{\rm AC}$   & 22:58:44.659 & +5:54:16.33 & AC & 0.4186   & 335.612  & 22.004 & 1.097 & 21.635& 0.825\\   
 \hline    
 \noalign{\smallskip}
     Pegasus~III &&&&&&&&&\\
\noalign{\smallskip}     
\hline 
     V1   & 22:24:18.312 & +5:22:17.78 & RRab  & 0.5537  & 357.560  & 22.58 & 1.18 & 22.13 &  0.96&$>-1.6/-1.7$\\ 
     V2$_{\rm RRL}$   & 22:24:01.932 & +5:25:45.52 & RRab & 0.6594    &296.771  & 22.33    & 1.26   & 21.80    & 1.16 &$>-$2.3\\ 
     V2$_{\rm AC}$   & 22:24:01.932 & +5:25:45.52 & AC & 0.3972    &357.531  & 22.37    & 1.03  & 21.85    & 0.84\\        
\hline    
\end{tabular}\\
\normalsize
\tablecomments{$^{*}$ From the CMDs analysis (see Sections~\ref{sec:v4_psc2} and \ref{sec:psc2_cmd}).}
\end{table*} 
\section{Distance to Psc~II and Peg~III}\label{sec:dist_psc2}                     We measured the distance to Psc~II using  V1, the bona-fide RRab star located within the galaxy r$_{h}$. The intensity-averaged $V$ magnitude of V1 is  $\langle V\rangle $ = 21.890 $\pm$ 0.038 mag (see Table~\ref{tab:psc2_var}), which we de-reddened using a standard extinction law A$_{V}$ = 3.1 $\times$ E($B-V$) and the reddening value from \citet{sef11}, E($B-V$)=0.056 $\pm$0.052 mag\footnote{An independent estimate of reddening can be obtained for V1 using the \citet{pie02} method for RRLs. From V1 we obtained a reddening value E($B-V$)=0.052$\pm$0.023 mag,  which agrees very well with the reddening from \citet{sef11}.}.
Similar to other nine  ultra-faint dwarfs
studied by our team (e.g. \citealt{dal06,gre08,cle12,gar13} and references therein) 
we adopted an absolute magnitude  M$_{V}$ = 0.54 $\pm$ 0.09 mag for RRLs with a metallicity of [Fe/H] = $-$1.5 dex (\citealt{cle03}) and the slope of the
luminosity-metallicity relation provided by \citet{cle03} and \citet{gra04}, namely $\frac{\Delta M_{V}}{\Delta[Fe/H]}$=0.214 $\pm$ 0.047 mag dex$^{-1}$. 
However,  later in the section we consider also more recent calibrations of the absolute magnitude of RRLs which are based on {\textit Gaia} parallaxes. 

 A metallicity for V1 to enter the RRL luminosity-metallicity relation, could in principle be obtained from the relation existing between the metal abundance ([Fe/H]), the  period and the  $\phi_{31}$,  parameter of the Fourier decomposition of the $V$-band light curve of RRab stars (\citealt{jur96}).  However, this relation cannot be used here because the $V$-band light curve of Psc~II V1 (as well as that of Peg~III V1)  does not satisfy the regularity conditions
(\citealt{jur96,cac05}) that allow for a reliable application of this method. For the metallicity of Psc~II we thus adopt
the value [Fe/H]$\sim$ $-$1.7 $\pm$ 0.1 dex which corresponds  to the metal abundance of the theoretical isochrones best fitting the galaxy CMD (see Fig.~\ref{fig:cmd_psc2_iso}, Section~\ref{sec:psc2_cmd}) dispite the spectroscopic metallicity estimated by \citet{kir15} is much lower ([Fe/H]=$-$2.45$\pm$0.07 dex).
The distance modulus we derived for Psc~II is then  $(m-M)_{0}$= 21.22 $\pm$ 0.14 mag, which corresponds to a distance of d = 175 $\pm$ 11 kpc. 
If we assume instead the \citet{kir15} spectroscopic metallicity, the distance modulus would be 0.16 mag longer, placing Psc~II $\sim$ 14 kpc farther than our previous estimate. 
 Within their respective errors, both these estimates are  consistent with \citet{san12}'s distance of 183 $\pm$ 15 kpc which is about in the middle of our  estimates.
\par
{\textit Gaia} second data release in spring 2018 
\citep[Gaia DR2,][]{gai18b}, published trigonometric parallaxes for more than a hundred thousand of sources classified as RRLs. 
We have adopted the  luminosity-metallicity relations from \citet{mur18}, which are calibrated on a sample of 381 Galactic RRLs with  \textit{Gaia} DR2 parallax measurements, to estimate the 
distance to Psc~II (and Peg~III). 
\textit{Gaia} parallaxes are known to be affected by a global zero-point offset with a mean value  $\Delta\varpi_{0}$ =  $-$0.03 mas in DR2 \citep{lin18,are18} and varying as a function of color, magnitude and sky position of the sources. For the RRLs 
the \textit{Gaia} DR2 parallax offset ranges from 
$-0.030$ to $-0.057$ mas \citep{mur18}. 
We summarise in Table~\ref{tab:dist} estimates of the distance modulus
 for Psc~II (upper and middle portions of the table) and Peg~III (lower portion), using the bona fide RRL in each galaxy and different choices for  the absolute magnitude of RRLs. In particular, the estimates based on {\textit Gaia} parallaxes  are taken from the relations in \citet{mur18} with parallax offset $\Delta\varpi_{0}=-0.03$, $-0.057$ (value fixed by \citealt{mur18} to infer the linear M$_V-$[Fe/H] relation using a subsample of 23 RRLs  with metallicity from high-resolution spectroscopy) and  $-0.062$ mas (this is the highest offset value considered by \citealt{mur18} 
 when fitting the linear M$_V-$[Fe/H] relation). Assuming  [Fe/H]=$-1.71 \pm 0.1$ dex for V1 in Psc~II we obtained $(m-M)_{0}$ = 21.12 $\pm$ 0.18 mag for both $\Delta\varpi_{0}=-0.057$ mas and $\Delta\varpi_{0}=-0.062$ mas. 
 These distance moduli place Psc~II closer than we find by adopting \citet{cle03} relation, but they are all still compatible  within  the errors. 
 There is excellent agreement  instead between the distance modulus from the \citet{cle03} relation and  the value $(m-M)_{0}$ = 21.25 $\pm$ 0.18 mag obtained using the \citet{mur18} relation with mean offset of $-0.03$ mas (see upper portion of Table~\ref{tab:dist}). 
Adopting the lower mean metallicity ([Fe/H]=$-2.45 \pm 0.07$ dex)  spectroscopically determined by \citet{kir15} (middle portion of Table~\ref{tab:dist}), we get identical results 
using \citet{cle03} relation and \citet{mur18} relation with parallax offset of $-0.057$ mas. These values are also in good agreement with \citealt{san12} estimation. Finally, distance moduli from \citet{mur18} relations with $\Delta\varpi_{0}=-0.03$ and $\Delta\varpi_{0}=-0.062$ place Psc~II, respectively, farther or closer than we find by adopting \citet{cle03} relation, but they are all still compatible  within  the errors. 
\par
Likewise, to estimate the distance to Peg~III,  
we have used the intensity-averaged $V$ magnitude of V1 ($\langle V\rangle $ = 22.126 $\pm$ 0.031 mag, see Table~\ref{tab:psc2_var}), the RRL with a firm  classification, which was  de-reddened using the reddening value E($B-V$)=0.126$\pm$0.003 mag from \citet{sef11} maps.   
 Then, following the procedure applied for  Psc~II, we  have adopted for the star the metal abundance [Fe/H] = $-$1.6 $\pm$ 0.20 dex, corresponding to the 
metallicity of the theoretical isochrones best fitting the Peg~III CMD 
(see right panel of Fig.~\ref{fig:cmd_psc2_iso}). 
This metallicity 
is significantly higher than the spectroscopic estimates by \citet{kim16} for the kinematically confirmed  members of Peg~III, which are as low as [Fe/H] = $-$2.55 $\pm$ 0.15 dex,  hence,  we have used  
both metallicity values to estimate the distance to Peg~III.
 We find the distance modulus  $(m-M)_{0}$ = 21.21 $\pm$ 0.23 mag (d= 174 $\pm$ 18 kpc) adopting for V1 the metal abundance from the isochrone fitting and 
 $(m-M)_{0}$ = 21.42 $\pm$ 0.19 mag (d= 192 $\pm$ 16 kpc) using the spectroscopic metal abundance from \citet{kim16}. 
 Both values  are shorter than the literature value of 215 $\pm$ 12 kpc, however,  the latter is consistent with the literature within the errors. 
 Finally, as done for Psc~II, we estimated the distance modulus of Peg~III from 
 \citet{mur18} luminosity-metallicity relations and three  values ($\Delta\varpi_{0}$ = $-0.03$, $-0.057$ and $-0.062$ mas, respectively) of the {\it Gaia} parallax offset (lower portion of Table~\ref{tab:dist}).
Again we find that all estimates are consistent within their admittedly large  errors, the \textit{Gaia} parallax zero-point offset  $-0.057$ and $-0.062$ mas put Peg~III closer than the value obtained from \citet{cle03} relation and, the  
 best agreement is found for $\Delta\varpi_{0}=-0.03$, 
 which leads to: 
 $(m-M)_{0}$ = 21.24 $\pm$ 0.25 mag. 

To ease the comparison with our previous studies of RRLs in UFDs, in the following sections we use for the distance moduli of Psc~II and Peg~III the estimates derived using  \citet{cle03} relation. 

\begin{table}[h]
\caption{Determinations  of the distance modulus for Psc~II (upper two  portions of the table) and Peg~III (lower two  portions).}
\centering
\footnotesize
\label{tab:dist}
\begin{tabular}{l c }
\hline
\hline
\noalign{\smallskip}
     Reference &  (m-M)$_{0}$\\
     & (mag)\\
\hline   
\noalign{\smallskip}
     Pisces~II \\
\noalign{\smallskip}
 \hline 
     assuming [Fe/H] = $-$1.7 $\pm$ 0.1 dex\\
\citet{cle03}& 21.22 $\pm$ 0.14\\
\citet[$\Delta\varpi_{0}=-0.03$]{mur18} & 21.25 $\pm$ 0.19\\
\citet[$\Delta\varpi_{0}=-0.057$]{mur18} & 21.12 $\pm$ 0.18\\
\citet[$\Delta\varpi_{0}=-0.062$]{mur18} & 21.12 $\pm$ 0.18\\
 \hline    
  assuming [Fe/H] = $-$2.45 $\pm$ 0.07 dex\\
 \citet{cle03}& 21.38 $\pm$ 0.14\\
\citet[$\Delta\varpi_{0}=-0.03$]{mur18} & 21.53 $\pm$ 0.18\\
\citet[$\Delta\varpi_{0}=-0.057$]{mur18} & 21.31 $\pm$ 0.16\\
\citet[$\Delta\varpi_{0}=-0.062$]{mur18} & 21.38 $\pm$ 0.17\\
 \hline
     Pegasus~III \\
\noalign{\smallskip}
\hline 
 assuming [Fe/H] =  $-$1.6 $\pm$ 0.2 dex\\
\citet{cle03}&21.21 $\pm$ 0.23\\ 
\citet[$\Delta\varpi_{0}=-0.03$]{mur18} & 21.24 $\pm$  0.25\\
\citet[$\Delta\varpi_{0}=-0.057$]{mur18} & 21.11 $\pm$ 0.24\\
\citet[$\Delta\varpi_{0}=-0.062$]{mur18} & 21.11 $\pm$ 0.25\\
\hline 
 assuming [Fe/H] =  $-$2.55 $\pm$ 0.15 dex\\
\citet{cle03} & 21.42 $\pm$ 0.19\\ 
\citet[$\Delta\varpi_{0}=-0.03$]{mur18} & 21.59 $\pm$ 0.21\\
\citet[$\Delta\varpi_{0}=-0.057$]{mur18} & 21.36 $\pm$ 0.20\\
\citet[$\Delta\varpi_{0}=-0.062$]{mur18} & 21.43 $\pm$ 0.21\\
\hline    
\end{tabular}\\
\normalsize
\end{table}
\section{Bailey diagram and Oosterhoff classification}\label{sec:oo_psc2}

\begin{figure*}[ht]
\centering
\includegraphics[trim= 40 150 0 100 clip, width=8.5cm]{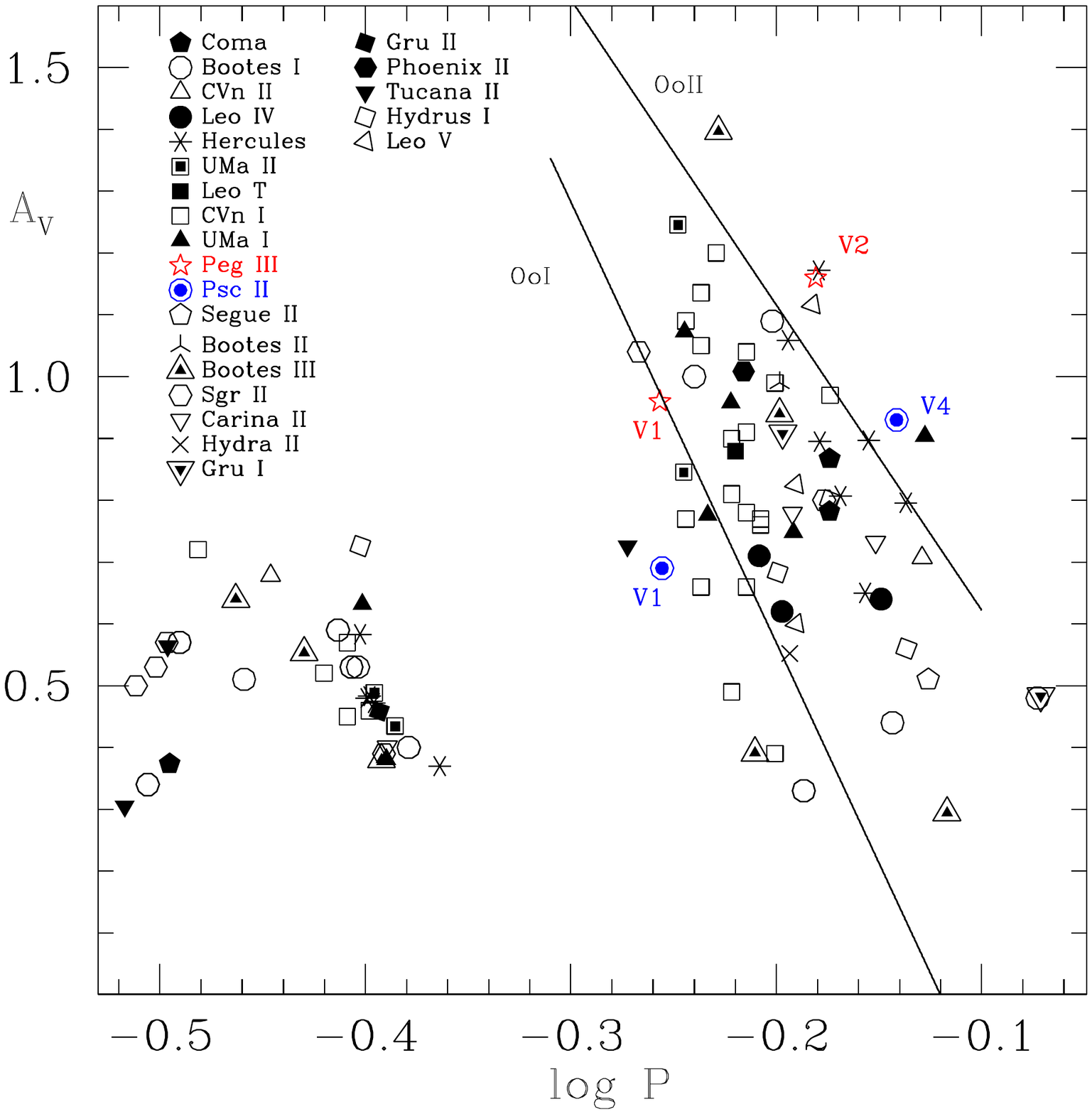}
\caption{Period-amplitude (Bailey) diagram: the bona fide RRab star inside the r$_{h}$ of Psc II (star V1) and the variable star with uncertain classification,  V4 (plotted adopting  the period as RRL, P $=$ 0.722 d) are shown as blue circles, while bona-fide RRab in Peg III, V1,  and  variable star with uncertain classification, V2 (assuming  the period as RRL, P $=$ 0.6594 d) are marked by red stars. Solid lines show the loci of Oo~I and Oo~II Galactic GCs according to \citet{cer00}. For comparison, we have also plotted the RRLs identified in other 21 MW UFDs (see text for details). We have transformed the SDSS $g$-band amplitudes, A$_{g}$, of the RRLs in Gru~I, Gru~II and Phoenix~II, to $V$-band amplitudes,  A$_{V}$, using a constant ratio, A$_{g}$/A$_{V}$ = 1.2, according to \citet[see their Figs.~11 and 12]{mar06}. The \textit{Gaia} $G$-band amplitudes of the RRLs in Bootes~I and III, Coma~Berenices,  Hydrus~I,  Tucana~II and UMa~II reported in 
\citet{viv20} were converted  to A$_{V}$ amplitudes  using  eq. 3 in \citet{clem16}.} 
\label{fig:bai_psc2_1}
\end{figure*}

We have used the $V$-band  amplitudes  and the periods of  the bona-fide RRab stars we have  identified in Psc~II and Peg~III, to plot the stars on the period-amplitude diagram (blue circles and red stars in  Fig.~\ref{fig:bai_psc2_1}, respectively)
and make a comparison with the RRLs in other MW UFDs. 

The RRLs detected in other 21 MW UFDs are also plotted in Fig.~\ref{fig:bai_psc2_1}
 using different symbols: Coma~Berenices \citep{mus09,viv20}, Bootes~I \citep{dal06,viv20}, Bootes~II and Bootes~III \citep{ses14,viv20}, Canes Venatici~I (\citealt{kue08}), Canes Venatici~II (\citealt{gre08}), Carina~II (\citealt{tor18}), Gru~I and Gru~II (\citealt{mar19}), Leo~IV (\citealt{mor09}), Leo~V \citep{med17}, Hercules \citep{mus12,gar18}, Hydra~II (\citealt{viv16}), Hydrus~I \citep{kop18,viv20}, Leo~T (\citealt{cle12}), Phoenix~II (\citealt{mar19}), Segue~II (\citealt{boe13}), Sagittarius~II (\citealt{joo19}), Tucana~II (\citealt{viv20}), Ursa Major~I (\citealt{gar13}) and Ursa Major~II \citep{dal12,viv20}. In the Bailey diagram fundamental mode and first overtone RRLs  lay  well separated. As 
expected for their periods
the RRLs we have identified in Psc~II and Peg~III fall in the region of the fundamental mode pulsators. 
In Fig.~\ref{fig:bai_psc2_1} we have highlighted with solid 
lines the loci of the two Oosterhoff types, Oo~I and  Oo~II, 
taken from \citet{cer00}. 
  The two bona-fide RRab stars in Psc~II and Peg~III
 fall on/close to the  Oo~I 
 line on the Bailey diagram.
 This is consistent with the metallicity we have inferred for these two RRLs from the isochrone fitting.  Conversely, Psc~II-V4 and Peg III-V2 both fall slightly beyond the Oo~II locus when adopting for them the  periods as RRLs.  
  It is clear from Fig.~\ref{fig:bai_psc2_1} that UFDs display a large dispersion in the period-amplitude diagram.  However,  what seems to be unique of Psc~II and Peg~III is the sharp separation between Oo~I and Oo~II types of the 2 RRLs each of these UFDs contains (under the assumption that Psc~II-V4 and Peg~III-V2 are indeed RRLs, see Sect.~\ref{sec:v4_psc2}). This seems to differ from what is generally found for other MW UFDs, whose RRLs tend to either lay preferentially between the two Oo lines, like in CVn~I (the system which more closely resembles the classical MW dSphs)  or to stay closer to just one of them, often the Oo~II locus (as in Hercules, Phoenix~II, Gru~I, Coma~Berenices and CVn~II).

\section{The intriguing cases of V4 in Psc~II and V2 in Peg~III}\label{sec:v4_psc2}

As anticipated in Section~\ref{sec:psc2_var}, V4 in Psc~II and V2 in Peg~III have an uncertain classification. The $B$ and $V$ data of V4 can as well be folded with P $\sim$ 0.72 d and the variable being an RRab star (as well as for V2 in Peg~III with P $\sim$ 0.66 d), or with P $\sim$ 0.42 d and the star being a first-overtone AC (as well as for V2 in Peg~III with P $\sim$ 0.40 d). ACs have been identified in a number of UFDs (i.e. CVn~I, \citealt{kue08}; Hercules, \citealt{mus12}; Leo~T, \citealt{cle12}; and Hydra~II, \citealt{viv16}). From an evolutionary point of view ACs can be associated with low metal abundance ([Fe/H]$\leqslant -1.4$ dex)
intermediate-age (1-2 Gyr) 
stellar populations, but neither Psc~II nor Peg~III show signs of intermediate age population in their CMDs (see Section~\ref{sec:psc2_cmd}).  However, ACs can also be formed by a binary evolution channel. In this case they would not trace a recent star formation episode, but would rather be the product  of binary interaction occurred about 1 Gyr ago \citep{gau17}.
This formation channel could justify the presence of ACs  in systems hosting  exclusively old stellar components like GGCs and most  of the MW UFDs.
We have performed a number of tests to investigate the nature of Psc~II-V4 and Peg~III-V2. 
 In particular, we have compared them with the $PL$ and $PW$ relations of ACs in the Large Magellanic Cloud (LMC) and,  with stellar evolutionary tracks for the typical mass and metallicity of ACs. These comparisons are presented in Appendix.~\ref{sec:appendixA}. They lead us to rule out a classification of V4 and V2 as 
ACs either belonging to the Psc~II and Peg~III UFDs or to the field around them.


\par
 To explore whether 
V4 in Psc~II and V2 in Peg~III  can be RRab stars with P=0.722 d  and 0.656 d,  respectively, we start first from their position on the 
period-amplitude diagram (Fig.~\ref{fig:bai_psc2_1}).
Both stars fall very close to the Oo~II locus and have a higher luminosity and lower metallicity than the RRab stars Psc~II-V1 and Peg~III-V1,  which lay  instead near the Oo~I locus.
Variable stars in Oo~II clusters have longer periods, higher luminosities and lower metallicities than those in the Oo~I clusters (see e.g. \citealt{cer00}, and references therein) and are supposed to be evolved off the Zero Age HB (ZAHB; see e.g. \citealt{lee90}). All the  properties of V4 and V2 are thus consistent with them being both RRab stars with Oo~II characteristics and evolved off the ZAHB. However, as we have already pointed out in Sect.~\ref{sec:oo_psc2}, 
although UFDs do show metallicity spreads as large as 1 dex among their members (see e.g. \citealt{sim19}, and references therein) and  large dispersions in the period-amplitude diagrams of their RRLs (Fig.~\ref{fig:bai_psc2_1}), what makes Psc II and Peg III a rather remarkable case is their capability to host RRLs with 
such distinct Oo properties.

\begin{figure}[ht]
\centering
\includegraphics[trim= 80 140 0 130 clip, width=9cm]{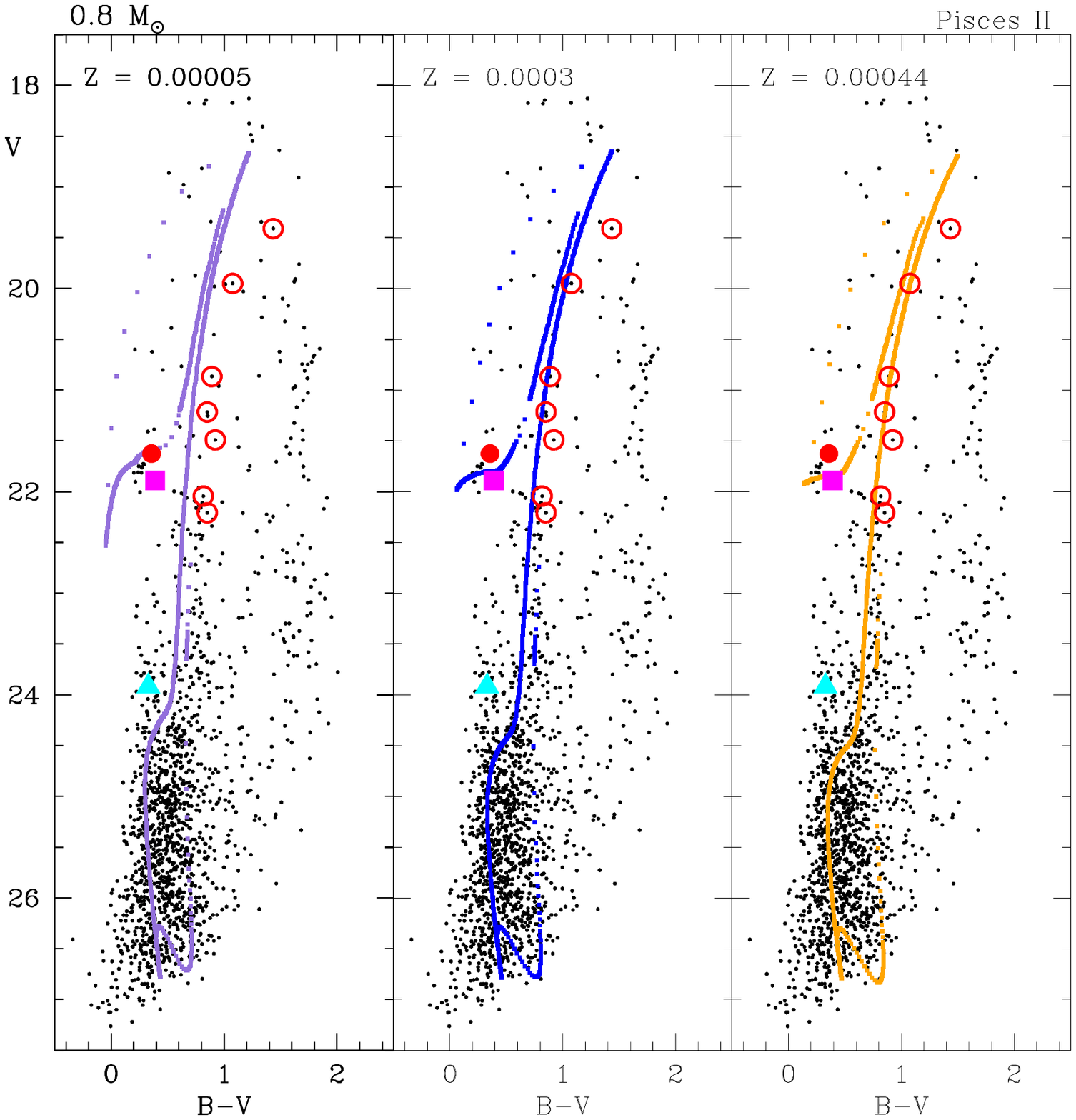}
\caption{Stellar  evolutionary  tracks  (BaSTI  web  interface) for 0.8  M$_{\bigodot}$ with  Z=0.00005 ([Fe/H]=$-$2.5 dex, violet line; left panel), Z=0.0003 ([Fe/H]=$-$1.8 dex, blue line; central panel) and Z=0.0044 ([Fe/H]=$-$1.5 dex, orange line; right panel). 
The tracks were corrected adopting the distance modulus of Psc~II derived from the RRab star V1 [$(m-M)_0$=21.22 mag] and the reddening E$(B-V)$=0.056 mag from 
\citet{sef11} maps. V1 is marked by a magenta filled square, star V4 by a red filled circle, the SX Phe star (V3) by a cyan filled triangle and the Psc~II spectroscopically confirmed members from \citet{kir15} by red open circles.}
\label{fig:track_psc2_new}
\end{figure}

While, it may prove  difficult to explain how Psc~II and Peg~III managed to produce RRLs of both Oosterhoff types (but see Section~\ref{sec:metalR_psc2}), a further option may be that Psc~II-V4  and Peg~III-V2 belong to structures/stellar systems projected in front of their hosts.  However, there is no sign of an  overdensity/structure surrounding V4 in the isodensity contour maps of 
Psc~II (see Fig.~\ref{fig:iso_psc2}). Moreover, the distance moduli inferred for Psc~II-V1 and V4 appear to be well consistent to each other, within the errors. 
This is illustrated in 
Fig.~\ref{fig:track_psc2_new}, which shows that stellar evolutionary tracks for 0.8 M$_{\bigodot}$ (a typical mass for RRLs) with metallicities corresponding to the metal abundance of V4 (Z=0.00005/[Fe/H]=$-$2.5 dex, violet line; left panel) and V1 (Z=0.0044/[Fe/H]=$-$1.5 dex, orange line; right panel), corrected for the distance modulus of Psc~II inferred from V1, well fit the luminosity of both V4 and V1, hence, supporting the conclusion 
 that they are both members of Psc~II. 
 
Nevertheless, it remains unclear how the two stars can belong to the same system although they are about 8 kpc apart in projected distance (if we trust the difference in distance moduli, see Section~\ref{sec:psc2_cmd}) and about 5 r$_{h}$ apart from the center of Psc~II. 


On the other hand, all variables we have identified in this study are outside the  r$_{h}$ of their hosts (except V1 in Psc~II),  therefore we cannot rule out they are field stars.
 Unfortunately,  they are fainter than the  \textit{Gaia} limit \citep[$G\leq$ 21 mag;][]{gai18b}, hence no proper motions are available for them from  \textit{Gaia} to check their membership to Psc~II and Peg~III.
However, we can evaluate the probability they are indeed members of these UFDs by calculating 
how many field RRLs can be expected in  
the MW halo  at the  distance of Psc~II and Peg~III  ($\sim$ 175 kpc). Assuming the radial density distribution of RRLs in the Galactic halo derived by \citet{med18},  within 145 kpc from the Galactic center, and considering that these  authors have identified 13 RRLs beyond 130 kpc over 120 deg$^{2}$, we found that the number of MW halo RRLs expected at $\sim$ 175 kpc in 0.15 deg$^{2}$ ($\sim$ the LBT FoV) is lower than 10$^{-9}$. 
Therefore, if V4 in Psc~II  and V2 in Peg~III are indeed RRLs, their membership to the corresponding host galaxy is very much plausible. 

\par 
To conclude, 
the most likely hypothesis is that Psc~II-V4 and Peg~III-V2 are RRab stars with Oo II characteristics belonging to respectively the Psc~II and Peg~III UFDs. 
However, it remains to clarify how Psc~II (and Peg~III) can host two stellar populations of different age and metallicity. May two separate star formation episodes have occurred in such  small galaxies? This is not quite a common scenario among the MW UFDs (\citealt{bla15,rom15} and references therein). 
  We also point out that  metallicity is not the only factor that may cause the difference in luminosity among the RRab stars in these galaxies. Other contributors   
 may be evolutionary effects,  photometric errors and, in particular, uneven sampling of the light curves. 
 Only the collection of new observations allowing us to definitively pin down the period and classification in type of V4 and V2 may help us  clarifying this issue. 
Adding new data would also help to better constrain the period, amplitude and mean magnitude of the other 
variable stars we have identified in these UFDs, hence 
strengthening their classification.

\section{Color-Magnitude Diagrams of Psc~II and Peg~III}\label{sec:psc2_cmd}
\begin{figure*}[ht]
 \centering
 \includegraphics[trim= 0 160 20 0 clip, width=9.5cm]{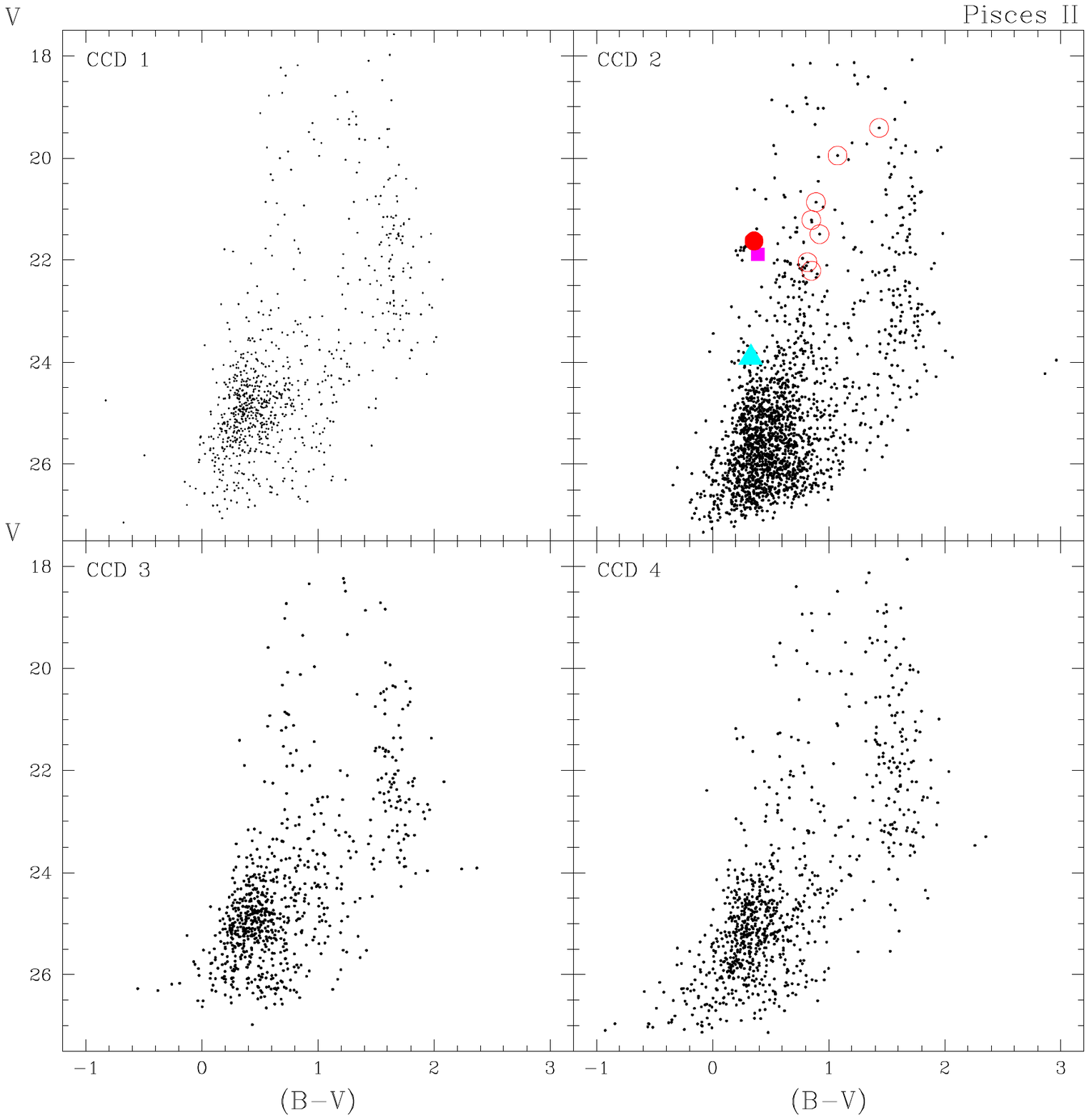}~\includegraphics[trim= 20 0 0 0 clip, width=8.5cm]{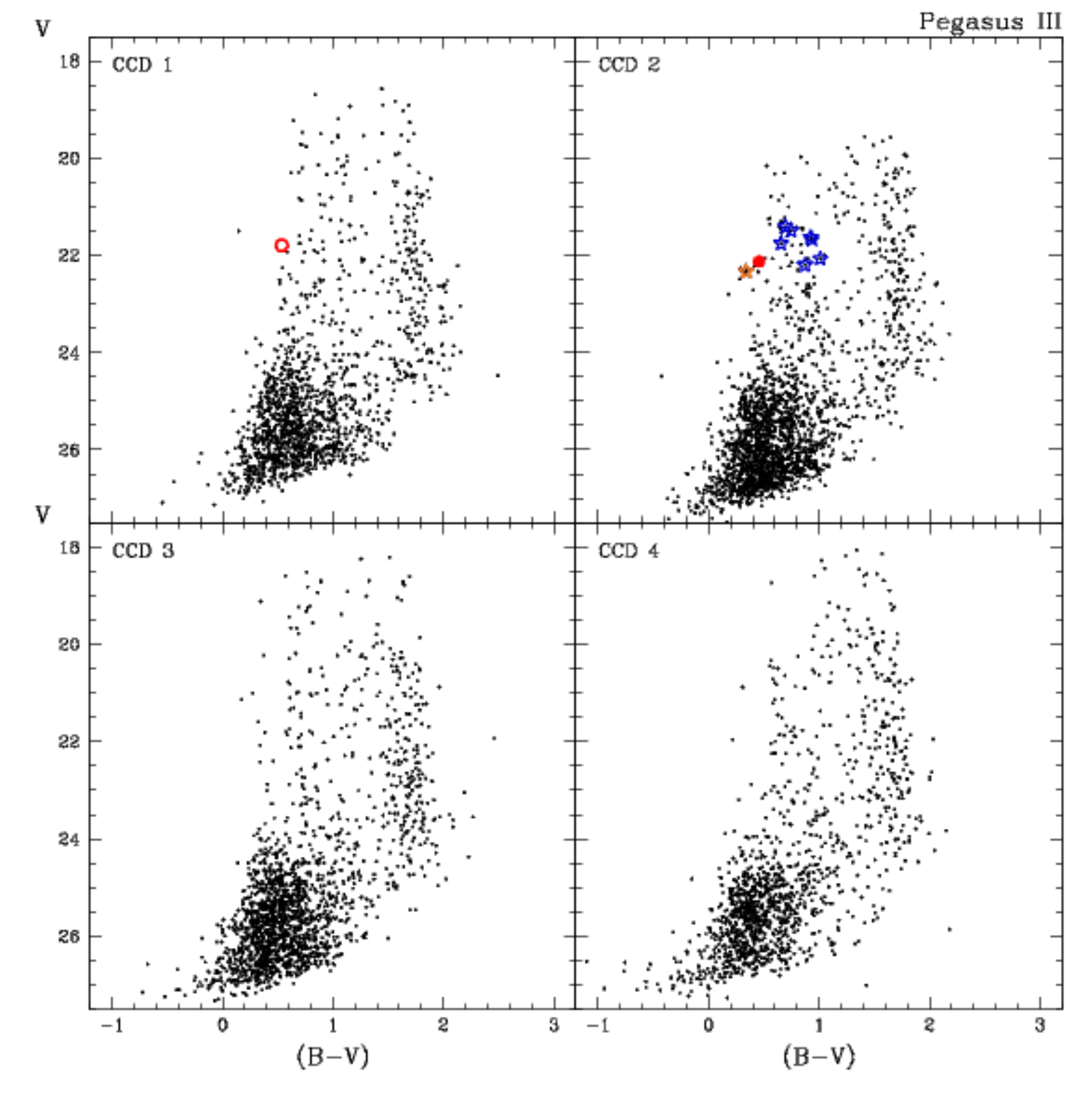}
 \caption{$V$, $B-V$ CMDs of Psc~II (on the left) and Peg~III (on the right) obtained from data corresponding to the four CCDs of the LBC mosaic, separately. 
Symbols and color-coding are the same as in Fig.~\ref{fig:map_psc2} 
for  Psc~II and Fig.~\ref{fig:map_peg3} for Peg~III.}
\label{fig:cmd_psc2_ccd}
\end{figure*}

\begin{figure*}[ht]
\centering
\includegraphics[trim= 0 0 0 0 clip, width=7.8cm]{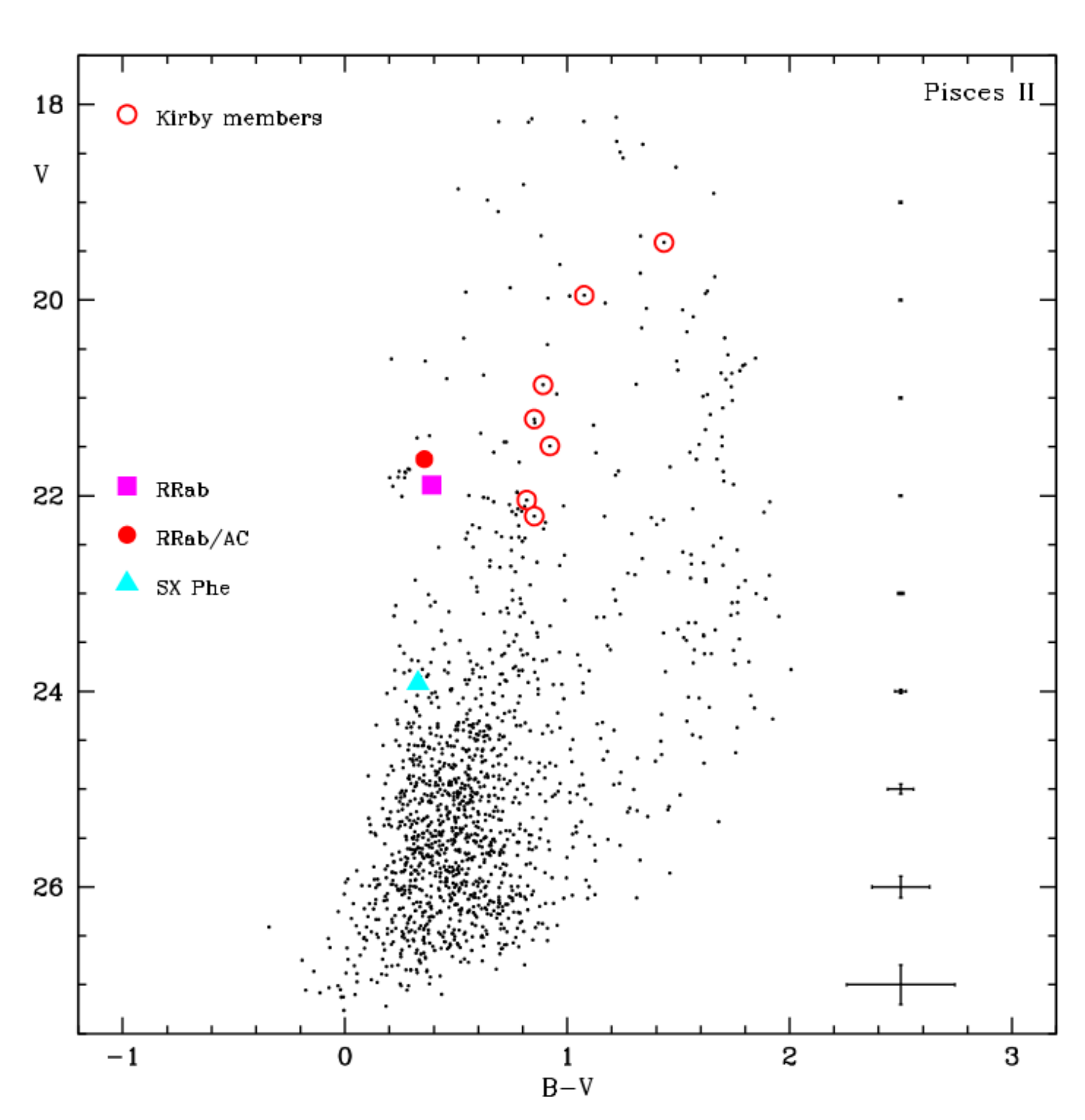}~\includegraphics[trim= 0 0 0 0 clip, width=7.8cm]{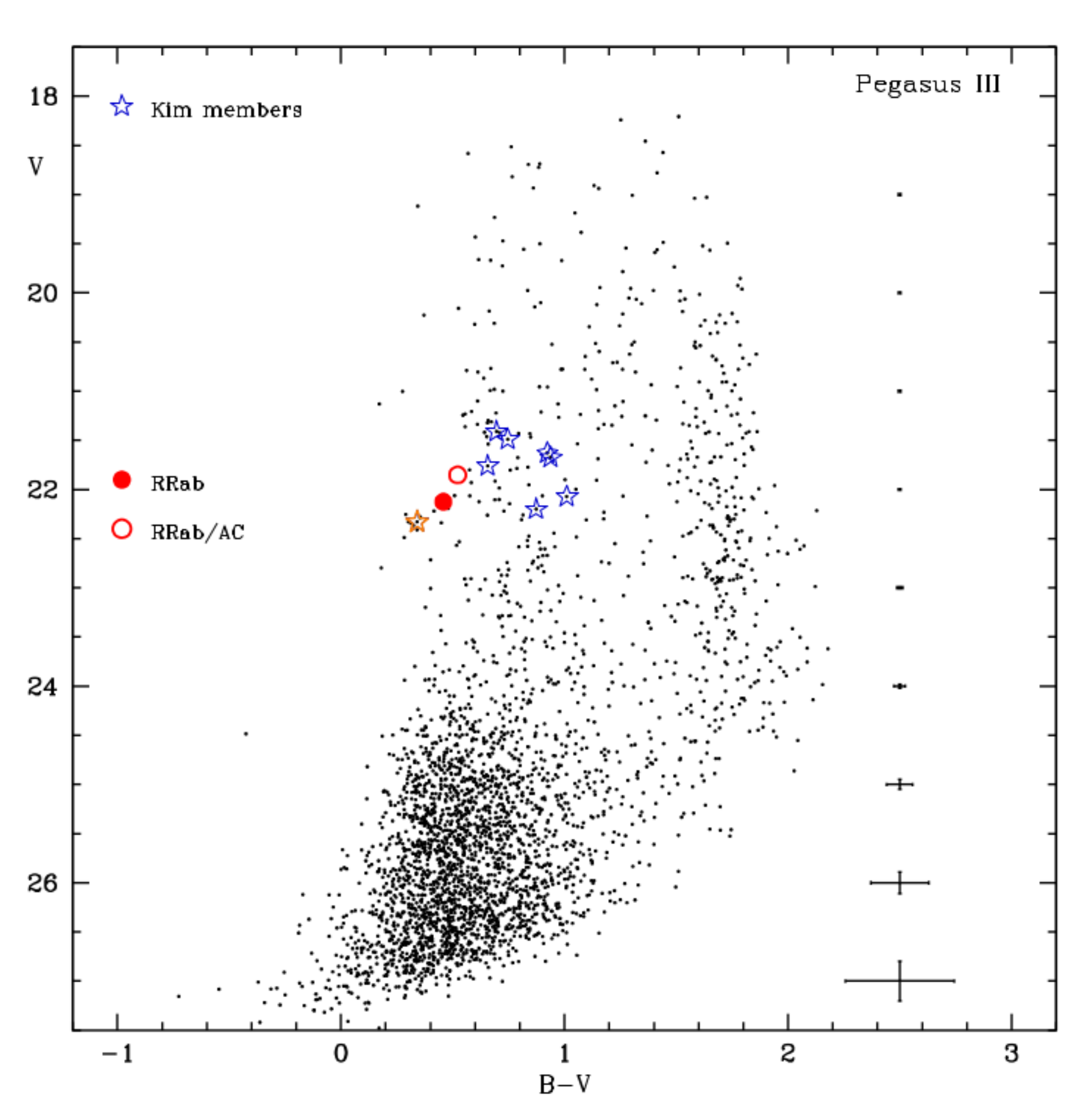}
\caption{$V$, $B-V$ CMDs showing stellar sources within five times the r$_{h}$ of  Psc~II  (left panel) and ten times the r$_{h}$ of Peg~III (right panel). 
Symbols and color-coding are the same as in Figs.~\ref{fig:map_psc2} and~\ref{fig:map_peg3}.}
\label{fig:cmd_psc2_5r}
\end{figure*}

The $V$, $B-V$ CMDs of Psc~II and Peg~III obtained in this study have already been introduced briefly in Sect.~\ref{sec:v4_psc2}. We now discuss how they were derived and show them side by side in  
Figs.~\ref{fig:cmd_psc2_ccd} and~\ref{fig:cmd_psc2_5r} to ease their comparison.

To clean the lists of  sources measured in each galaxy we have used the quality information provided by  
the \texttt{DAOPHOT} quality image  parameters $\chi$ and $Sharpness$. Only the stellar detections satisfying the following photometric quality  criteria: $-$0.4 $\leq  Sharpness \leq$ 0.4, and $\chi<$ 2.2, in both $B$ and $V$ images were retained in the CMDs. This allowed us to reduce the contamination by background galaxies. In Fig.~\ref{fig:cmd_psc2_ccd} we have plotted all the stellar detections in the LBT FoV for each galaxy, selected according to the above $\chi$ and $Sharpness$ values and  separated according to 
the four CCDs of the LBC mosaic. 
 We note that Milky Way stars dominate the CMDs in Fig.~\ref{fig:cmd_psc2_ccd} for colors redder than $B-V \sim$1.6 mag.
In order to better identify stars belonging to the Psc~II galaxy 
we have cross-matched our catalog against the \citet{kir15} list of spectroscopically confirmed members of Psc~II. All the 7 RGB stars which  are Psc~II members according to \citet{kir15} have a counterpart in our catalog.  They are  marked by red,  empty circles in the left panels of Fig.~\ref{fig:cmd_psc2_ccd}. 
A magenta square marks V1, the bona-fide RRab star, a red filled circle V4, the variable with less certain classification, both plotted according to their intensity-averaged mean magnitudes computed along the pulsation cycle. A cyan filled triangle shows the SX Phe star (V3). Similarly, in the right panels  
of Fig~\ref{fig:cmd_psc2_ccd}, the two variables identified in Peg~III, V1 and V2, are marked by filled and open red circles, respectively, and plotted according to their intensity-averaged mean magnitudes and colors, the spectroscopically confirmed members by  \citet{kim16} are marked by blue stars, and  an orange star symbol marks the source whose membership is uncertain. Since the ambiguous member lies on the HB region, \citet{kim16} speculated that it might be an RRL. Based on our checks of the variability indices in $B$ and $V$ and the visual inspection of the light curves 
we rule out that this star varies.

\begin{figure*}[ht]
\centering
\includegraphics[trim= 0 10 0 0 clip, width=8cm]{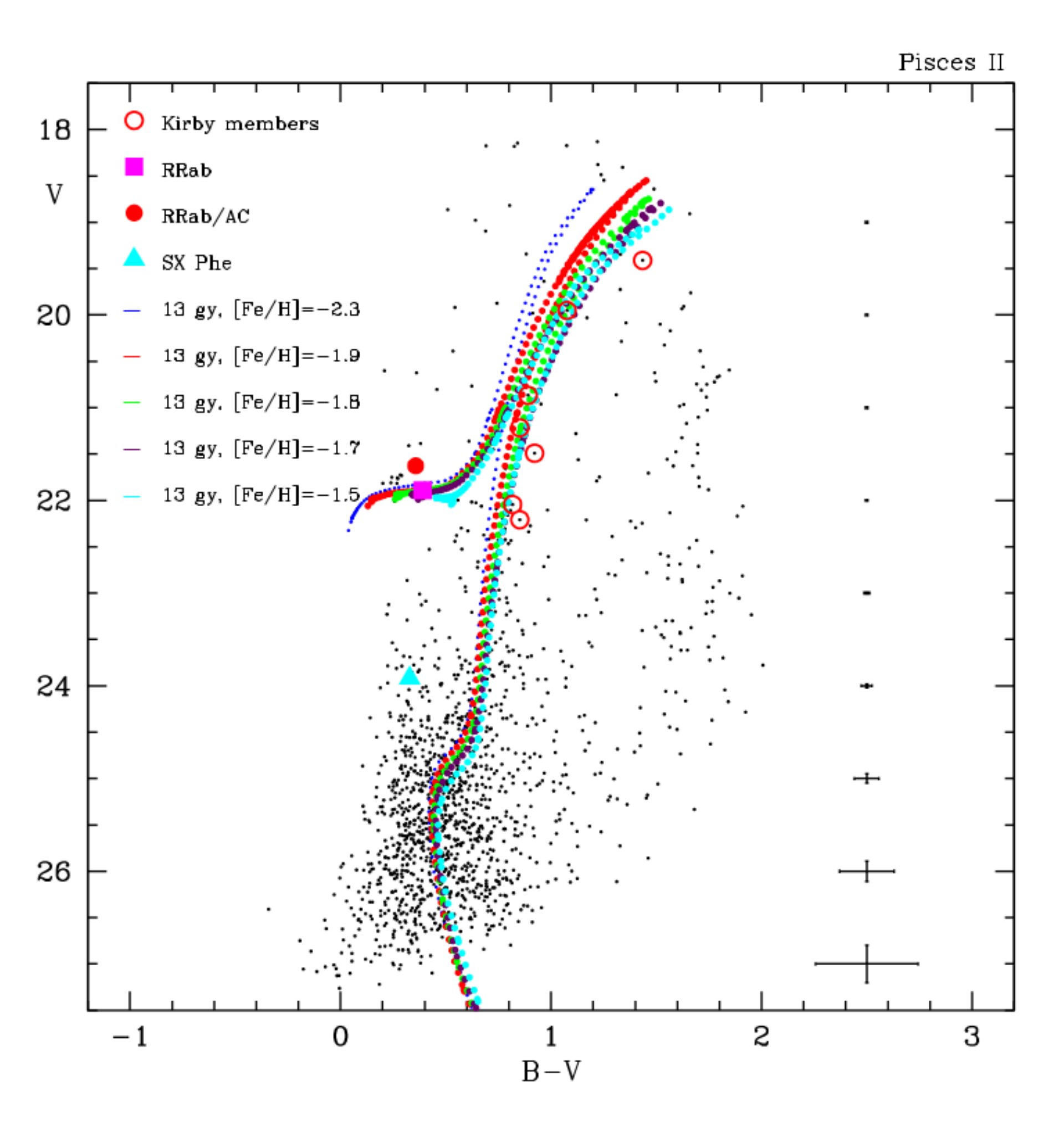}~\includegraphics[trim= 0 0 0 0 clip, width=8cm]{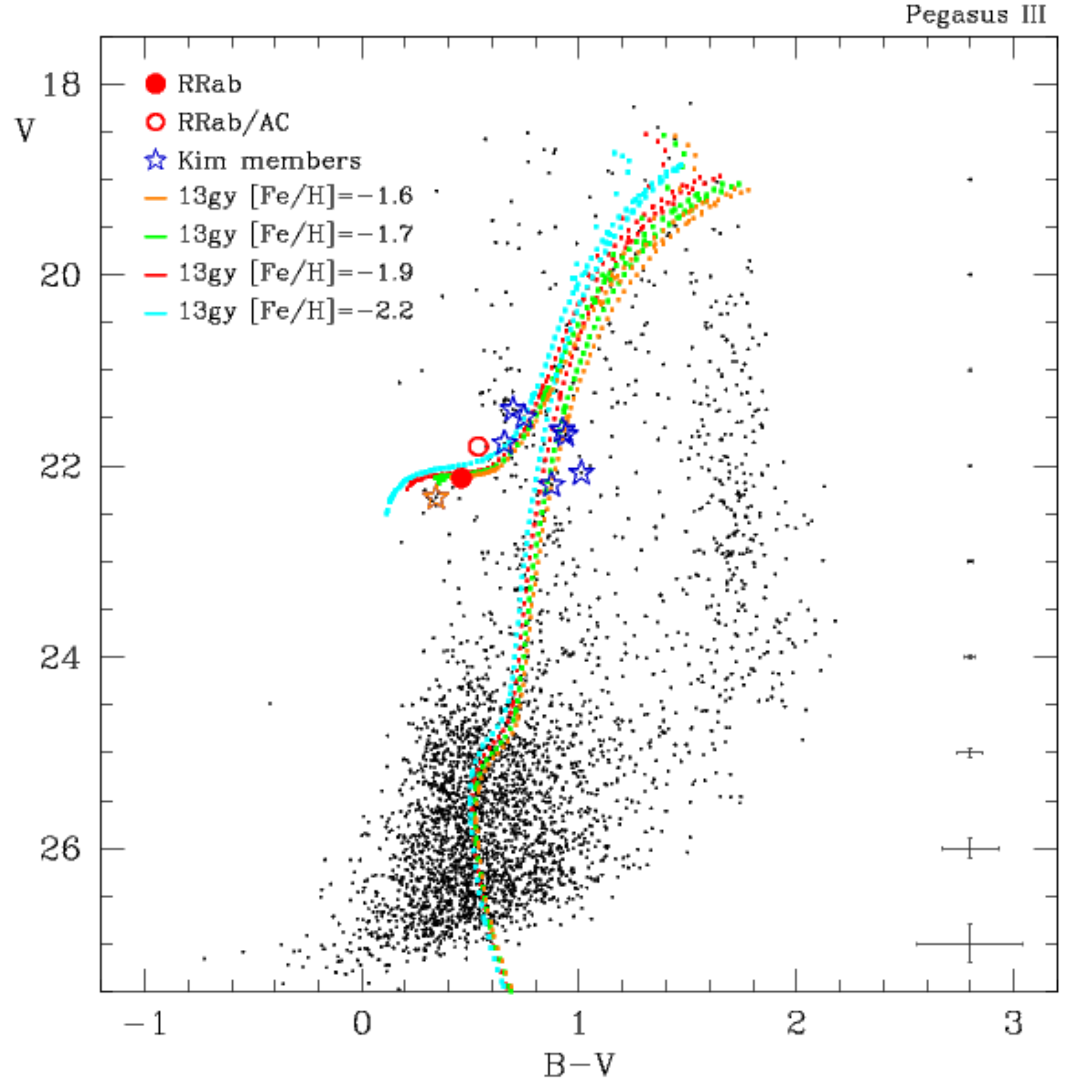}
\caption{\textit{Left}: Same as in the left panel of Fig.~\ref{fig:cmd_psc2_5r}, but with dotted lines showing the PARSEC stellar isochrones (\citealt{bre12}) for an age of 13 Gyr and metallicities of: [Fe/H] = $-$2.3 dex, blue; [Fe/H] = $-$1.9 dex, orange; [Fe/H] = $-$1.8 dex, green; [Fe/H] = $-$1.7 dex, purple; and [Fe/H] = $-$1.5 dex, cyan. The isochrones were corrected for a distance modulus of 21.22 mag, as derived from the bona fide RRab star (V1) in Psc~II, and for a foreground reddening of E($B-V$) = 0.056 mag, according to the \citet{sef11} maps.
\textit{Right}: Same as in the right panel of Fig.~\ref{fig:cmd_psc2_5r}, dotted lines in different colors are the PARSEC stellar isochrones for an age of 13 Gyr and metallicities of [Fe/H]= $-$1.6, $-$1.7, $-$1.9 and $-$2.2 dex; they were corrected for a distance modulus of 21.21 mag, as inferred from the bona fide RRab star (V1) in Peg~III and for a foreground reddening of E($B-V$) = 0.13 mag (\citealt{sef11}). Symbols and color-coding are the same as in Figs.~\ref{fig:map_psc2} and~\ref{fig:map_peg3}.}

\label{fig:cmd_psc2_iso}
\end{figure*}

Our CMDs of Psc~II and Peg~III are rather deep, reaching $V\sim$ 26.5 mag, hence they allow us to trace for each galaxy the main sequence (MS) below the turn-off (TO) point, which  is located approximately between 25 and 25.5 mag in $V$,  according to the theoretical isochrones overlaid on the CMDs (see Fig.~\ref{fig:cmd_psc2_iso}, later in this section).

As shown by the left panels  of Fig.~\ref{fig:cmd_psc2_ccd}, the Psc~II galaxy is entirely  contained in CCD2, the central CCD of the LBC configuration, where are located all the spectroscopically confirmed  members and the variable stars we have identified in Psc~II.  
Only in CCD2 is a hint of the galaxy HB and RGB recognisable (for the latter, the eye is guided by the spectroscopic members).
The CMD of Psc~II within the half-light radius is very poorly populated; only V1 and one of the 7 spectroscopic members are within the galaxy $r_h$, the other six confirmed members are  located in the region between 1 and 4 times the galaxy r$_{h}$. We have tried to build a final CMD reasonably well populated, which contains all spectroscopic members and the variable stars in the field of Psc~II.
This is presented in the left panel of Fig.~\ref{fig:cmd_psc2_5r},  
which shows   
all stellar sources located within 5 times the r$_{h}$ of the galaxy (all the CMDs shown in the remaining of this section correspond 
to sources   within 5 r$_{h}$).
\par
The Peg~III UFD is also very small in size (r$_{h}= 53 \pm$ 1.4 pc; \citealt{kim16}).
As shown by  Fig~\ref{fig:map_peg3} the bona-fide RRab star (V1) and the variable of uncertain classification are located respectively within 5 and 10 times the galaxy r$_{h}$, whereas the spectroscopic members are all within 3r$_{h}$. 
Since both spectroscopic members and variable stars are rather far away from the center of Peg~III, even more than for Psc~II, the CMD we show in the right panel of Fig.~\ref{fig:cmd_psc2_5r}
contains all stars within 10 times the r$_{h}$ of Peg~III. 
Furthemore,  unlike Psc~II, Peg~III seems to be not entirely contained in CCD2,  as one  would expect given its small size.  V1 along with the center coordinates of Peg~III are  within CCD2,  whereas V2 
is within CCD1.
The most recognisable features in the CMDs of Psc~II and Peg~III  are: (i) the HB at $V\sim$ 22 mag (as also inferred from the mean $V$ magnitude of the bona-fide RRab star in each UFD) populated by a few stars close to the red and blue edges of the RRL  instability strip and roughly ending at $B-V\sim$ 0.2 mag; 
(ii) the RGB between ($B-V$)$\sim$0.7 and $\sim$1.4 mag, reaching as bright as $V\sim$ 19.5$-$19 mag, which is  disentangled thanks to the RGB member stars by \citet{kir15} for Psc~II and  \citet{kim16} for Peg~III. Additionally, in the CMD of Peg~III a possible hint of  asymptotic giant branch (AGB) or red horizontal branch (RHB) population can be recognised, as also suggested  by \citet{kim16}, thanks to three kinematic members brighter than the HB and bluer than the RGB (colors 0.6 $<$ ($B-V$ ) $<$ 0.8 mag).
These features are indicative of a stellar population older than 10 Gyr in both Psc~II and Peg~III, in agreement with the typical ages of the stellar components observed in other MW UFDs. In order to estimate the age and metallicity of the Psc~II and Peg~III dominant stellar populations, we have over-plotted on the  CMDs the PARSEC stellar isochrones from \citet{bre12}, available on the CMD 2.9 web interface\footnote {\url{http://stev.oapd.inaf.it/cmd}}. The left panel of Fig.~\ref{fig:cmd_psc2_iso} shows the CMD of Psc~II in a region within 5 times the galaxy r$_h$ with overlaid the PARSEC isochrones for a fixed age of 13 Gyr (the best-fit we have obtained) and 5  different values of metallicity: [Fe/H] = $-$2.3, $-$1.9, $-$1.8, $-$1.7 and $-$1.5 dex. The isochrones were adjusted to the distance modulus of Psc~II (21.22 mag, as inferred from V1, the bona fide RRab star within the galaxy  r$_h$) and a reddening value of E($B-V$)= 0.056 mag from the  \citet{sef11} maps. The comparison with the stellar isochrones shows that Psc~II hosts a dominant ancient stellar population ($\sim$ 13 Gyr) more metal-rich than [Fe/H]$\sim -$1.8$\pm$0.1 dex, assuming that the bona fide RRab star (magenta square) traces the HB of Psc~II and the spectroscopically confirmed members (red empty circles) of \citet{kir15} trace the galaxy RGB. In the right panel of Fig.~\ref{fig:cmd_psc2_iso} we show the CMD of Peg~III within 10 $r_h$, with  overlaid PARSEC  isochrones for a fixed age of 13 Gyr and  metallicities ranging from [Fe/H] = $-$2.2 to $-$1.6 dex. The isochrones were corrected for a reddening E($B-V$)= 0.13 mag from the \citet{sef11} maps  and a distance modulus of 21.21 $\pm$ 0.32 mag, as inferred from the bona-fide RRab star in Peg~III (V1). The isochrones which better reproduce the RGB and the HB level along with the position of  Peg~III-V1 
are those for metallicities between $-$1.7 and $-$1.6 dex, however the isochrone with [Fe/H] = $-$1.6 dex does not reproduce the overall extension of the HB, which, instead, is well covered by the [Fe/H] = $-$1.7 dex isochrone. The metallicity we find from the  isochrone fitting is much higher than the value inferred by \citet{kim16} applying the same technique.
 Using isochrones from 
\citet{dot08}, \citet{kim16}  suggest as best-fit  isochrone for Peg~III the one with an age of 13.5 Gyr and metallicity  [Fe/H] = $-$2.5 dex. However, we think that the isochrone set with higher metallicity ([Fe/H] = $-$1.5 dex, right panel of fig. 2 in \citealt{kim16}), better reproduces also the CMD by \citet{kim16}.

In summary, our results from the CMD isochrone fitting 
suggest that both Peg~III and Psc~II host a 
stellar population more metal-rich than derived from  spectroscopically confirmed members or by isochrone fitting in  previous works.\footnote{ We have tried to investigate what may be causing these discrepant results by exploring different sets of isochrones including also alpha-enhancement, since UFDs are expected to have enhancement of alpha-elements. In particular, in Appendix.~\ref{sec:appendixB}, we have considered \citet{dot08} isochrones (available at:  http://stellar.dartmouth.edu/models/).} 
 Among the MW UFDs, only Leo~T \citep{cle12} and Ursa Major~II \citep{dal12} have a similarly high metallicity. 
  On the other hand, the comparison with isochrones of different metallicity presented here, the very low metallicity measured for the spectroscopically  confirmed members 
and the position on the period-amplitude diagram also  suggest a plausible scenario where   V4 in Psc~II  and V2 in Peg~III are very metal-poor   ([Fe/H]$>-$2.4 dex) RRab  members and tracers of a  very  metal poor component of these UFDs.  
\begin{figure*}[ht]
\centering
\includegraphics[trim= 0 160 0 0 clip, width=9.cm]{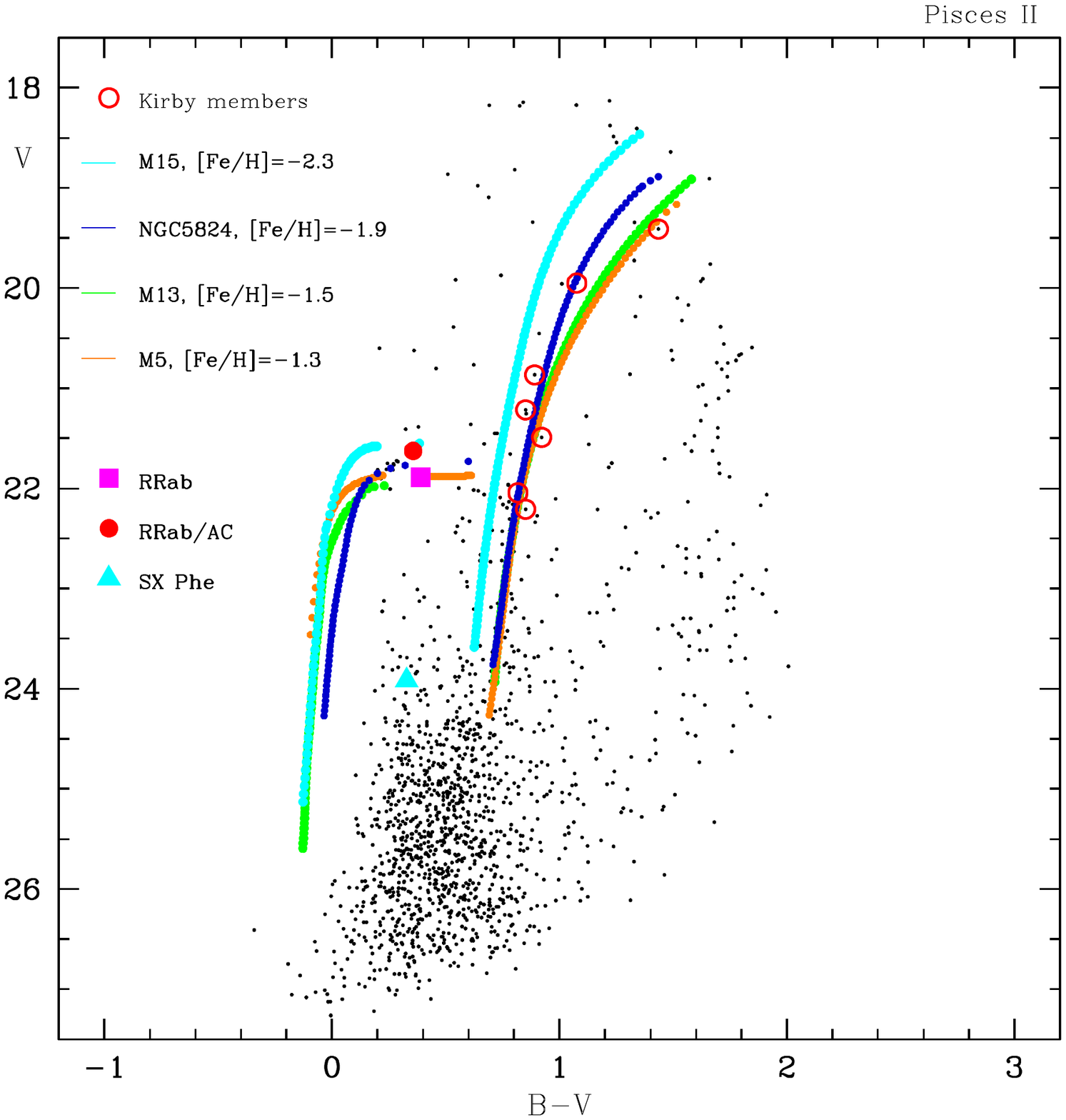}~\includegraphics[trim= 0 0 0 0 clip, width=8.cm]{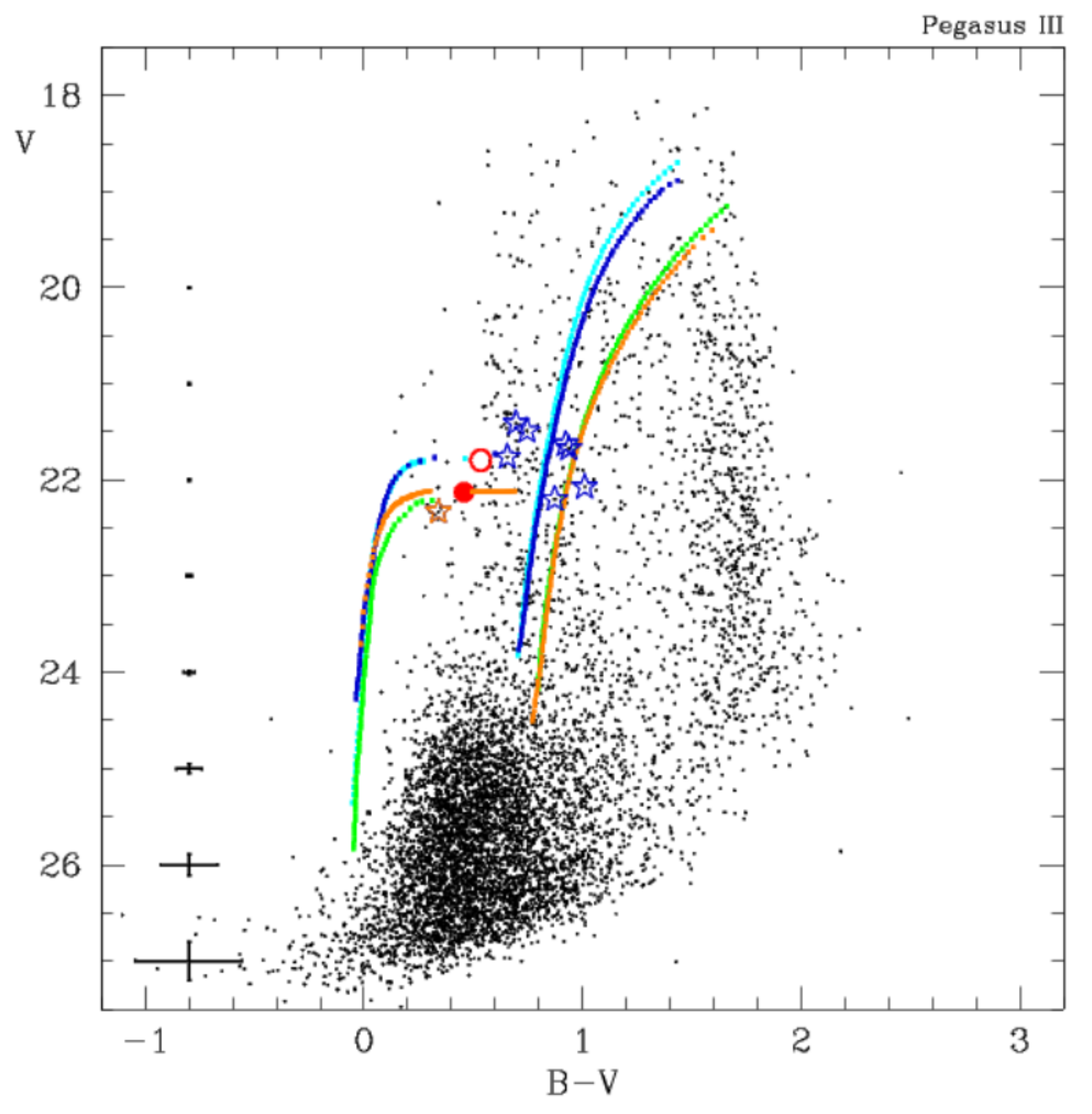}
\caption{Same as in  Fig.~\ref{fig:cmd_psc2_5r} but with solid lines showing the ridge lines of the Galactic GCs: M15 (cyan), NGC 5824 (blue), M13 (green), and M5 (orange).}
\label{fig:cmd_psc2_rl}
\end{figure*}
Indeed, as we have shown in Sect.~\ref{sec:dist_psc2}, if we adopt for V4 in Psc~II the metallicity derived by \citet[Fe/H=$-$2.45$\pm$0.07 dex]{kir15} the inferred distance modulus is $(m-M)_{0}$= 21.12 $\pm$ 0.20 mag, which would be    consistent,  within the errors, with the distance modulus inferred from the more metal rich RRab star V1  (21.22 $\pm$ 0.14 mag).
In the same vein, 
if V2 is 
a very metal poor ([Fe/H] = $-$2.54 $\pm$ 0.79 dex) RRab star belonging to Peg~III,  its distance modulus (21.07 $\pm$0.42 mag) would be  consistent,  within the errors, with the distance modulus
inferred from the more metal rich RRab star V1 in Peg~III (21.21 $\pm$ 0.32 mag).

Such very metal poor stellar components as those possibly 
traced by V4 in Psc~II and V2 in Peg~III,   should be  associated with steep RGBs in the CMDs.
Unfortunately there is no clear sign of such a  metal-poor RGB, steeper than the metal-rich RGB associated with V1, in the CMD of Psc~II (see left panels of Figs.~\ref{fig:cmd_psc2_ccd} and ~\ref{fig:cmd_psc2_5r}).  
However, the bright portion of the Psc~II CMD is so scarcely populated, that even the metal-rich RGB would hardly be 
recognised if there weren't the spectroscopically confirmed members to guide the eye.
Conversely, a  metal-poor RGB, steeper than the metal-rich RGB associated with V1 can be  recognizable in the CMDs of Peg~III (see right panels of  
 Figs.~\ref{fig:cmd_psc2_ccd} and ~\ref{fig:cmd_psc2_5r}). 
 \par 
As both the PARSEC isochrones (CMD 2.9 web interface) and the isochrones by \citet[BaSTI web interface\footnote{ \url{http://basti.oa-teramo.inaf.it/BASTI/WEB_TOOLS/IM_HTML/index.html}}]{pie04} do not reach metallicities lower than [Fe/H]$\sim -$2.2/$-$2.3 dex, 
as a last check, we compared the  CMDs of Psc~II and Peg~III to the mean ridge lines of metal poor Galactic GCs.
The left panel of Fig.~\ref{fig:cmd_psc2_rl} shows the mean ridge lines of the Galactic GCs used to fit  the RGB (with V1 and \citealt{kir15} spectroscopic members as references) and the HB (with V1 and V4 as references) of the two different populations possibly observed in Psc~II. Specifically, we used the fiducial tracks published by \citet{pio02} for M15 and NGC 5824, which have metallicities respectively of [Fe/H] = $-$2.4 and $-$1.9 dex, to fit the metal-poor component, and the fiducial lines of M13 and M5, with metallicities respectively of [Fe/H] = $-$1.5 and $-$1.3 dex, to fit the metal-rich component in Psc~II. The ridge lines were corrected according to the proper reddening and distance modulus of each GC (\citealt{har96}\footnote{Revision of December 2010}): E($B-V$) = 0.1 mag and $(m-M)_{V}$ =15.39 mag for M15, E($B-V$)=0.13 mag and $(m-M)_{V}$ =17.94 mag for NGC 5824, E($B-V$) = 0.02 mag and $(m-M)_{V}$ = 14.33 mag for M13, E($B-V$)= 0.03 mag and $(m-M)_{V}$ = 14.46 mag for M3 and then further shifted to the distance modulus (21.22 $\pm$ 0.14 mag) and reddening E($B-V$)= 0.056 $\pm$ 0.052 mag (\citealt{sef11}) of Psc~II. The ridge line of NGC5824, corresponding to a metallicity of [Fe/H] = $-$1.9 dex, well traces the RGB and the position of V4, while M5 and M13 well reproduce the RGB and the position of V1. 
We conclude that Psc~II might host two old stellar components, a metal-poor component ([Fe/H]$>-2.3$ dex), traced by V4, and a higher metallicity component traced by V1 ([Fe/H]$>-1.6/1.7$ dex). Totally similar results are found by over-plotting the ridge lines of the same GCs to the CMD of Peg~III. This is shown in the right panel of Fig.~\ref{fig:cmd_psc2_rl} where we have assumed for Peg~III the distance modulus inferred from the RRab star (V1). As done for Psc~II, to fit the HB defined by V1 and the RGB of the metal-rich component we used the fiducial lines of M13 and M5, while the RGB and HB of the metal-poor component are well traced by the fiducial lines of M15, corresponding to a metallicity of [Fe/H] = $-$2.4 dex. 
The fit of the Peg~III CMD with the GC ridge lines  confirms the results and the metallicities suggested by the fit with the theoretical isochrones, thus leading to conclude that 
also Peg~III might host two old stellar components with different metallicity, a higher metallicity component traced by V1 and a metal poor component traced by V2.
\section{On metal retention in Psc~II and Peg~III}\label{sec:metalR_psc2}
In this section we discuss how  small systems like Psc~II and  Peg~III may have managed to form  stars with such a large dispersion on the metallicity  distribution as \citet{kir15}  measured in Psc~II ($\sigma$[Fe/H]=0.48$\pm^{0.70}_{0.29}$);  a spread which is confirmed by our isochrone fitting of the   CMDs (see Sect.~\ref{sec:psc2_cmd}) and the properties of the variable stars we have identified in these two galaxies.
\par
The present-day stellar masses ($M_{*}$) of Psc~II and Peg~III can
be derived by their absolute  $V$-band magnitudes (Table~\ref{tab:psc2_main}) and
from the stellar mass-to-light ratio of a low-metallicity stellar population,
which at an age of $\sim$10 Gyr is of the order of $\sim 2.73$ \citep{bru03} 
assuming a \citet{sal55} stellar initial mass function (IMF).
With these assumptions and for a solar $V$-magnitude $M_{V}=4.83$ mag, one obtains $M_{*} = 10000~M_{\odot}$ and 
$M_{*} = 5300~M_{\odot}$ for Psc~II and for Peg~III, respectively.
These stellar mass values lie at the lower end of the values found for Local Group dwarf galaxies \citep[e. g.][]{kir13}.

These assumptions imply that each of them must have hosted $\sim~10^2$ massive stars
(adopting the stellar mass of Peg~III as reference value and again assuming a Salpeter IMF)  
and possibly a number of core-collapse supernovae (SNe) 
of the same order of magnitude must have exploded throughout their history.
Assuming that each supernova (SN) has produced $\sim 0.1$ $M_{\odot}$ of Fe, 
this also implies that these stars must have synthesised and ejected in total $\sim$ 10 $M_{\odot}$ in the form of Fe. 
Present-day stars show a spread in [Fe/H] of 0.3-0.6 dex, much larger than the one shown 
by stellar systems of comparable mass, i.e. GCs, which is commonly of the order of 0.1 dex \citep[except a few cases such as $\omega$ Cen,][]{ren15}.
Similar large metallicity spreads (up to 0.5 dex) are common in other MW UFDs \citep[e. g.][]{seg07}. 

This is likely to indicate that, at variance with most GCs and as other UFDs, 
Psc~II and Peg~III were able to retain a significant amount of the metals produced by SNe and incorporate it
into new stars.
This fact seems to contrast with the idea that, despite the small number of massive stars hosted by each of the two 
systems, the total amount of energy released by such stars in both the form 
of stellar winds and SNe exceed the binding energy of the gas. 
An order-of-magnitude estimate of the binding energy can be performed as
\begin{equation}
E_b = \frac{G~M_{gas}~M}{r_h}, 
\end{equation}
where $M_{gas}$ is the gas mass present in the system when most of its stars formed, $G$ is the gravitational
constant and $M$ and $r_h$ are the total mass and half-light radius, respectively. 
Assuming  that during star formation $M_{gas}~\sim~M_{*}$ and the data of Table~\ref{tab:psc2_main} for the effective radii $r_h$ and for the dark matter
halo mass of the two galaxies,
one finds that in both cases one single SN would be enough to completely expel all the natal gas.

However, the requirement for the SN energy release to exceed the binding energy is a necessary
but not sufficient condition to expel all the metal-rich gas from a bound system. 
One of the reasons is that
the conversion efficiency of the energy injected by SNe into kinetic energy of the ISM is generally well below unity, since
a considerable fraction is lost via radiative losses \citep[e.g.][]{mor02}. 
Another important argument is related to the spatial distribution of massive stars.
In a low-density system such a dwarf galaxy, massive stars are likely to have formed in isolated associations, 
scattered within the entire extent of host galaxy, with important consequences 
on the effects of their energetic feedback. 
Previous works have shown that if isolated SN explosions occur at random times,  
it is difficult to reach the conditions for the gas to be heated at high temperatures 
to generate a steady wind \citep{nat13}. This is mostly due to the difficulty of 
SNe remnants to overlap and heat the gas at sufficient temperatures to achieve an outflow
(typically $T>10^6~K$). 
Massive stars are likely to have originated grouped in some OB associations, 
but in that case the feedback sources would be 
even more isolated throughout the volume over which the gas is distributed.
In principle, the simultaneous action of the continuous winds blown by massive stars in the pre-SN phase
produces interstellar bubbles which could merge in a short time
and lead to a large porosity of the hot gas. 
This process has turned out to be efficient in particular in stellar clusters,
where the relative distances between OB associations are of the order of $\le 1 pc$ and
these can act simultaneously to rapidly achieve a large thermalization efficiency
of the ISM \citep{cal15,yad17,sil18}.
However, even in the case of sources acting simultaneously,
not always an outflow will be driven. In fact, off-center feedback sources might 
drive inward-propagating shocks which compress 
the gas in the innermost regions, with the effect of enhancing the radiative losses \citep{mor02,rom19}. \\
It is worth noting that even in conditions in which a massive outflow
cannnot originate, the expulsion of metals produced by SNe is facilitated with respect to the cold gas, 
as metals are injected at high velocity and in a hot phase,
in particular in the outskirts of a galaxy or when the density of the surrounding medium is low. 
However, given the low metallicty of the two systems, the retention of 
a very small amount of Fe is enough to originate the [Fe/H]
spread. In fact, considering an initially metal-free system and an 
average present-day stellar metallicity of [Fe/H]=$-2.3$ 
in Peg~III, it is sufficient to retain and incorporate into stars $0.05$ $M_{\odot}$ of Fe, 
which corresponds to 0.5 \% of the total amount of Fe produced by Type II SNe, and presumably an even smaller fraction to account for a $0.6$ dex spread in Fe.\\
It is worth noting that this estimate does not include the Fe released by Type Ia SNe, which explode on longer timescales
than type II SNe and which in general release a larger amount of Fe \citep{mag17}.\\
The data set collected in this work does not allow us to assess the precise duration of the star formation episode
which gave place to the stars of Peg~III and Psc~II. To derive this quantity, the collection of
a larger sample of stars with measured
Fe abundances is required, which will allow one to compute a metallicity distribution function (MDF).
This quantity can be studied 
by means of chemical evolution models and, as it is known to depend on the infall time scale,
can provide strong contraints on the duration of the star formation history \citep[e.g.][]{vin19}. 
\section{Isodensity contour maps}\label{sec:iso_psc2}
\begin{figure*}[ht]
\includegraphics[trim= 0 0 0 0 clip, width=6cm]{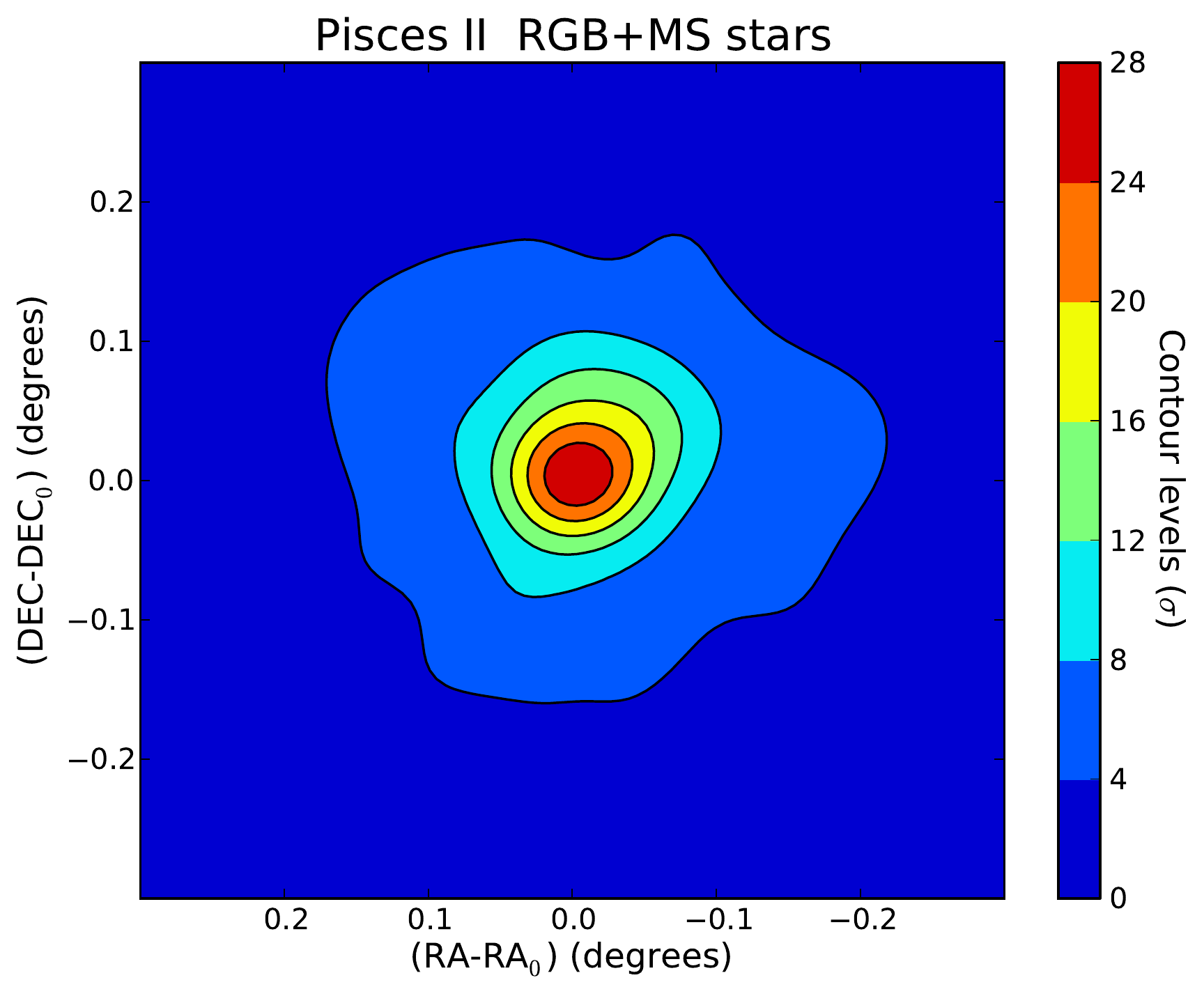}
\includegraphics[trim= 0 0 0 0 clip, width=6cm]{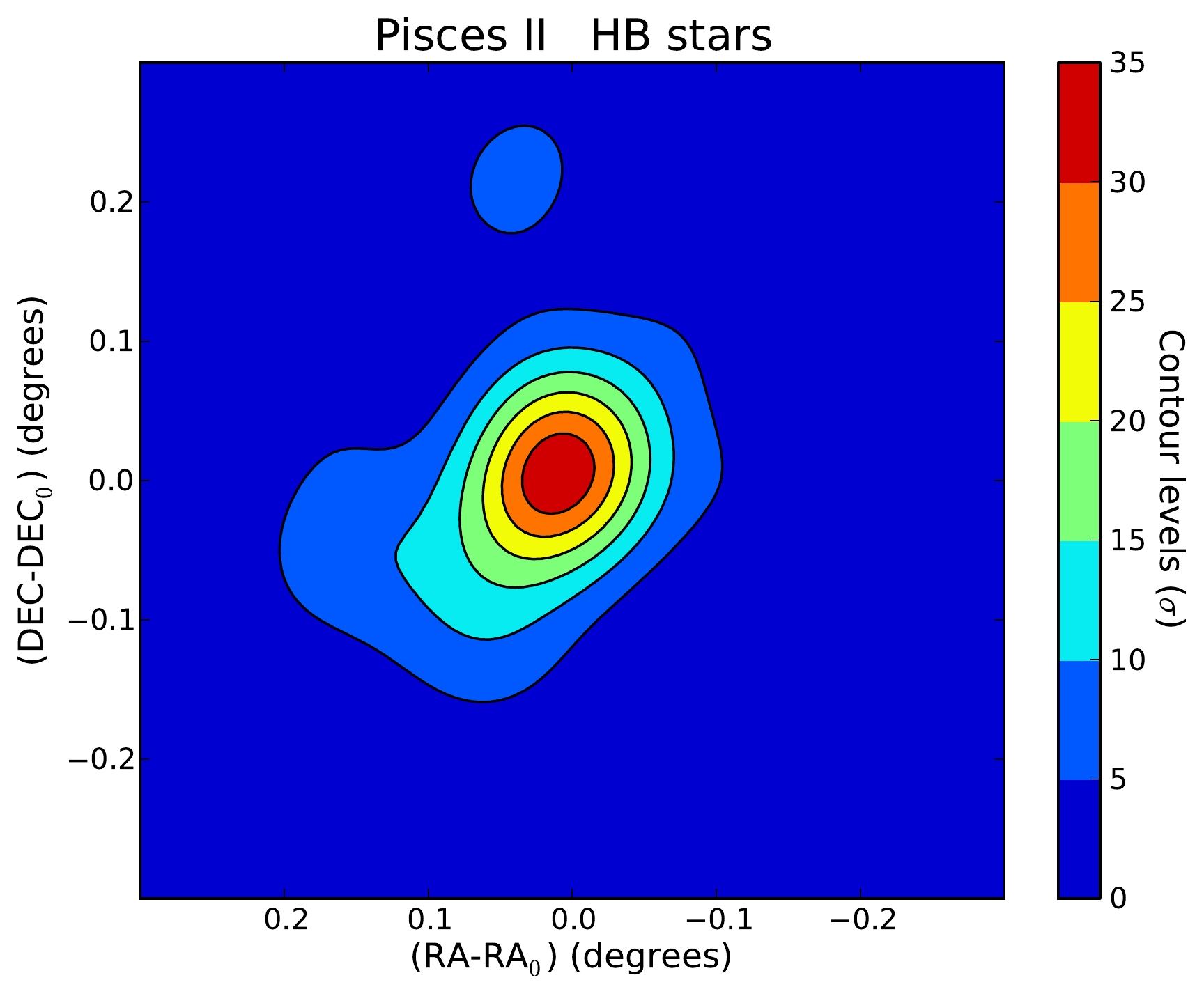}~\includegraphics[trim= 0 0 0 0 clip, width=6cm]{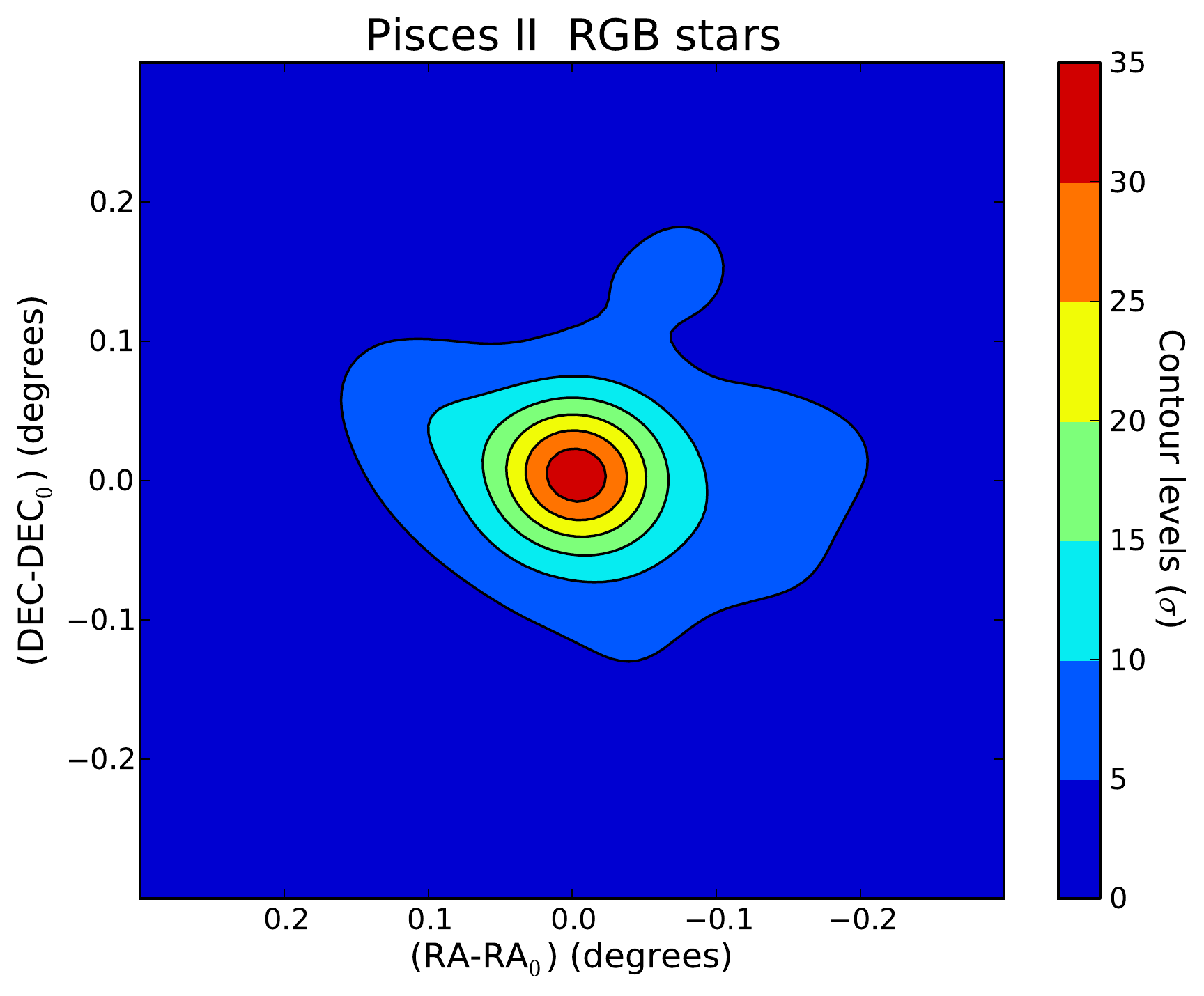}
\caption{Isodensity contour maps of Psc~II obtained with a bin size of 0.23$^{\prime}$. \textit{Left}: Isodensity map of MS and RGB stars, selected with the guidance of  the theoretical isochrones. \textit{Center}: Same as in the left panel but for HB stars.  \textit{Right}: Same as in the left and middle panels but for RGB stars only. Peg~III is located in the bottom-right direction on these maps. 
 The colorbar provides  the number of $\sigma$s above the background represented by each isodensity contour level.
 The isodensity contour levels  in the maps of Psc~II vary from 4 to 35 $\sigma$ above the background.}
\label{fig:iso_psc2}
\end{figure*}

\begin{figure*}[ht]
\centering
\includegraphics[trim= 0 0 0 0 clip, width=6.7cm]{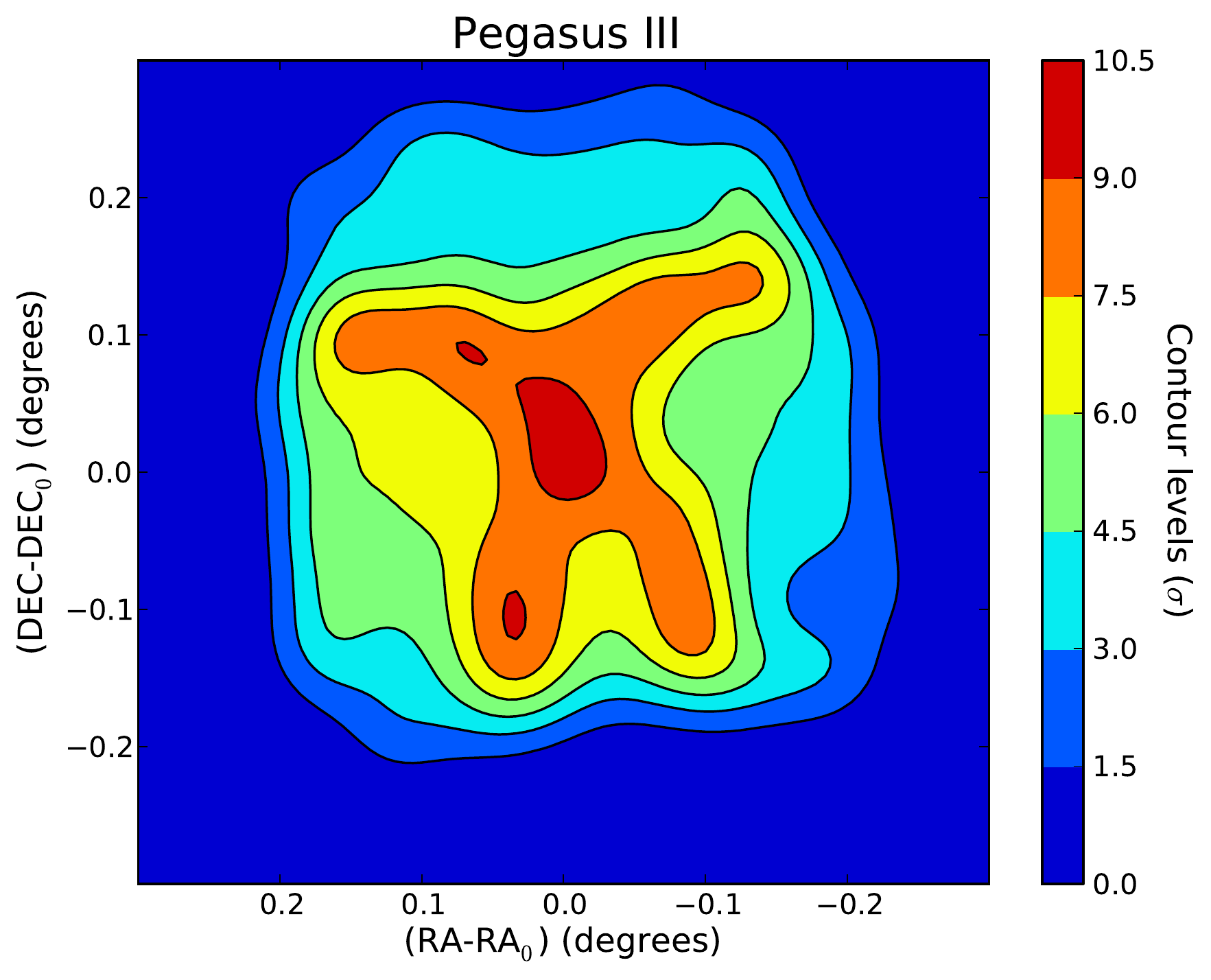}~\includegraphics[trim= 0 0 0 0 clip, width=6.7cm]{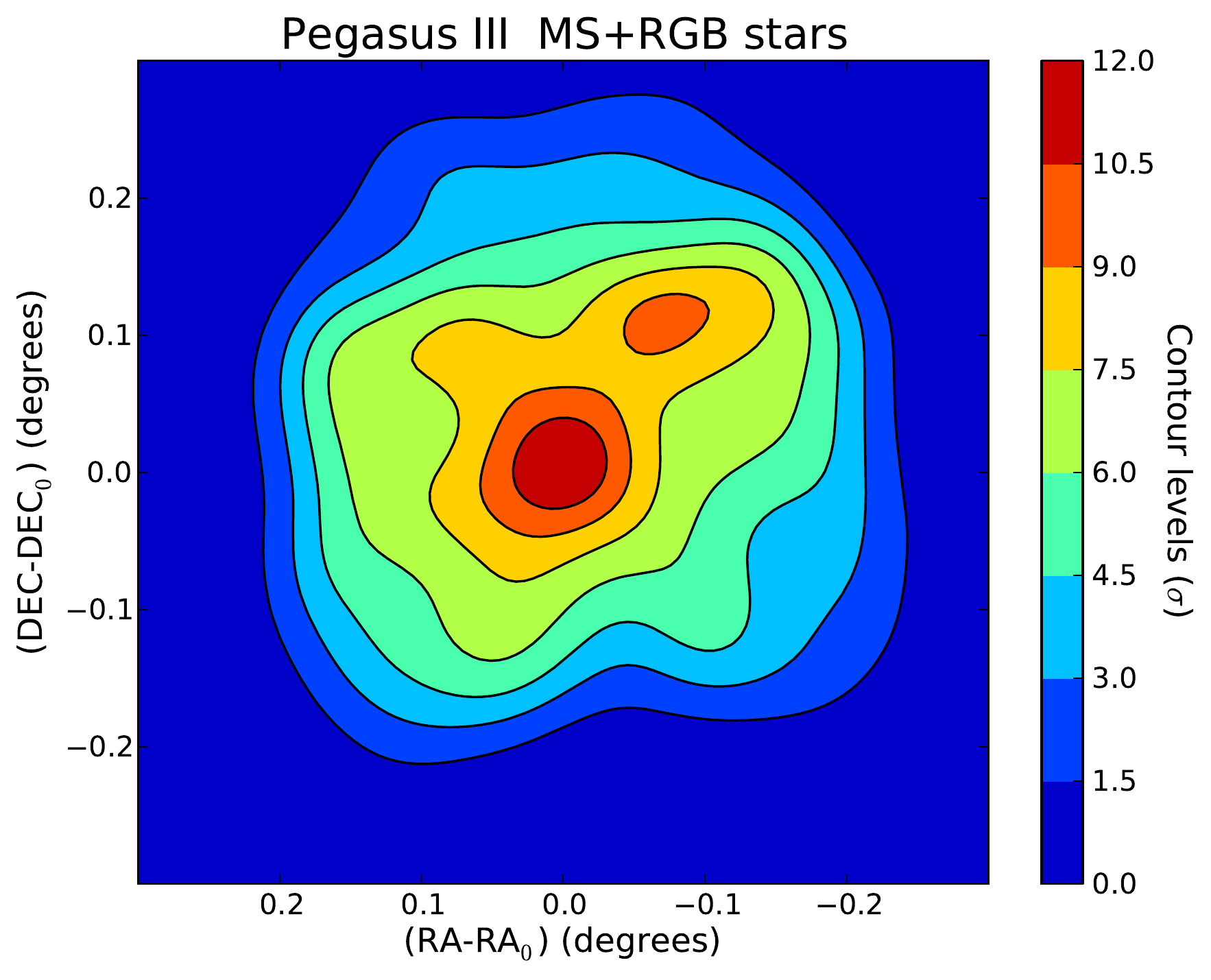}
\caption{Isodensity contour maps of Peg~III obtained with a bin size of 0.23$^{\prime}$. \textit{Left}: Isodensity map from all stars in the photometric catalog of Peg~III selected by $\chi$ and Sharpness parameters. \textit{Right  panel}: Same as in the left panel but only for MS and RGB stars  selected with the guidance of the theoretical isochrones. 
Psc~II is located in the top-left direction on these maps. The colorbar provides  the number of $\sigma$s above the background represented by each isodensity contour level. The 
isodensity contour levels in the maps of Peg~III vary from 1.5 to 12 $\sigma$ above the background.}
\label{fig:iso_peg3}
\end{figure*}

Figure~\ref{fig:iso_psc2} shows the isodensity contour maps obtained from all stars in our photometric catalog of Psc~II, 
selected by the $\chi$ and $Sharpness$ conditions described in Section~\ref{sec:psc2_cmd} and using the same limits as for the galaxy CMD.  
The selected stars have been binned adopting a bin size of 0.23 arcmin. 
 The left  panel 
shows the isodensity contour map obtained selecting only MS and RGB stars (1387 sources), the center panel only HB stars (24 sources) and the right panel only  RGB stars (309 sources).  RGB, MS and HB stars were selected with the guidance of the theoretical isochrones we have overlaid on the galaxy CMD and, to minimize the contamination by field sources   we considered only stars  within 0.1 mag from the best fitting isochrones.  To quantify a possible contamination of our 
subsamples by MW stars, on the assumption that the MW stars are uniformly distributed over the whole LBT FoV, we have  counted the number of sources in 1 arcmin$^2$ 
regions at about 10 arcmin in distance from the centre of Psc~II ($\gtrsim$ 10 $r_{h}$). 
This number corresponds to 6 sources, on average, which  
we can assume are all 
 background/foreground objects. Since within an area of 1 arcmin$^2$ 
centered on Psc~II we find 54 sources, we can conclude that
at most $\sim$ 11 $\%$ of the sources 
in our subsamples may be MW contaminants. 
On the other hand, since the MW stars dominate the reddest part of the CMD ($B-V \geqslant$ 1.6 mag, see Fig~\ref{fig:cmd_psc2_5r}), and barely overlaps with 
the  Psc~II RGB, MS and HB stars, we can 
safely assume that background/foreground  contamination is almost negligible in the 
 isodensity contour maps 
 in Figure~\ref{fig:iso_psc2}. 
 
All  maps in  Figure~\ref{fig:iso_psc2} 
 confirm the presence of a stellar overdensity at their center, corresponding to the position of the Psc~II galaxy. 
This overdensity 
 is clearly seen also in the isodensities of the RGB only and HB only stars. 
 The isodensity maps of Psc~II do not show an irregular shape as one would expect if the galaxy was undergoing tidal disruption or was interacting with Peg~III. For sake of completeness,  Peg~III is located in the bottom-right direction on  these maps.

Figure~\ref{fig:iso_peg3} shows the isodensity contour maps of
Peg~III obtained using the same bin size and selection of the sources by $\chi$ and $Sharpness$, as adopted for Psc~II. The left panel corresponds to sources on the whole CMD    
(6557 stars) whereas the right panel corresponds to only  MS+RGB stars  selected by overlying the theoretical isochrones on the CMD of Peg~III (1125 sources).
Both isodensity maps show an overdensity at the center, corresponding to the Peg~III galaxy. 
Other more extended overdensities are visible in the maps in Figure~\ref{fig:iso_peg3}.
They are not real structures but rather artefacts caused by regions with deeper exposures  (see Fig.~\ref{fig:map_peg3}) due to the LBT rotation discussed in  Sect.~\ref{sec:psc_obs}. The main overdensity at the center of the  map on the left panel of Figure~\ref{fig:iso_peg3} is clearly seen also in the isodensity map of the RGB and MS stars, while the fictitious overdensities have now almost disappeared. Similarly to Psc~II, the isodensity maps of Peg~III do not show an irregular shape. 
The regular shape of the Psc~II and Peg~III isodensity maps 
leads us 
to rule out the existence of a clear link or stellar stream between these two  UFDs.

\section{Conclusions}\label{sec:conc_psc2}

Using $B$, $V$ time series photometry obtained with the  LBCs at the LBT we have performed the first study of the variable stars in the Psc~II and Peg~III UFDs and derived the main properties of their  resolved stellar populations. 
We have focused on the comparison of  the Psc~II and Peg~III properties,  in order to  investigate the existence of a physical connection between these two galaxies which have  a spatial separation of only 43 $\pm$ 19 kpc, as it was suggested in previous studies \citep{kim15, kim16}. 
For this comparison we have used as main tools (i) the properties of the RRLs identified in the two  systems, (ii) the features of the observed CMDs and, (iii) the density contour maps.

In Psc~II we have identified 4 variable stars (upper portion of Table~\ref{tab:psc2_var}): an RRab star (V1) with P $\sim$ 0.56 d, an SX Phoenicis star (V3; P $\sim$ 0.06 d) both lying within the r$_h$ of Psc~II, and  a third variable, V4, with uncertain classification and about 0.25 mag brighter than V1, which is outside the galaxy's r$_{h}$.
A forth source within the r$_{h}$, V2, shows variability only in the $B$ band with P$\sim$0.05 d and was classified as candidate SX Phoenicis star.  
 The period and amplitude of the light variation place V1 
on the Oo~I locus of the period–amplitude diagram. 
From the mean magnitude of V1 we measured a distance modulus: 21.22 $\pm$ 0.14 mag, that places Psc~II at 175  $\pm$ 11 kpc from us, in 
agreement,  within the errors, with previous literature estimates and at the same distance of Peg~III. The period of V4 is uncertain. The source could either be an RRab star with P = 0.72 d or a first overtone AC with P = 0.42 d. However, from  the comparison with theoretical isochrones overlaid on the galaxy CMD, and the $PL$ and $PW$ relations for ACs and CCs in the LMC, 
we conclude that V4 is likely a metal poor ([Fe/H] $\gtrsim$ $-$2.4 dex) RRab star with  Oo~II pulsation characteristics belonging to Psc~II. The CMD suggests the presence in Psc~II of a dominant old stellar population (t $>$ 10 Gyr) with metallicity [Fe/H] $> -$1.8 dex, along with, possibly, a minor, more metal poor component traced by Psc~II-V4.\\

\begin{figure}[hbp]
\centering
\includegraphics[trim= 50 160 0 90 clip, width=9.0cm]{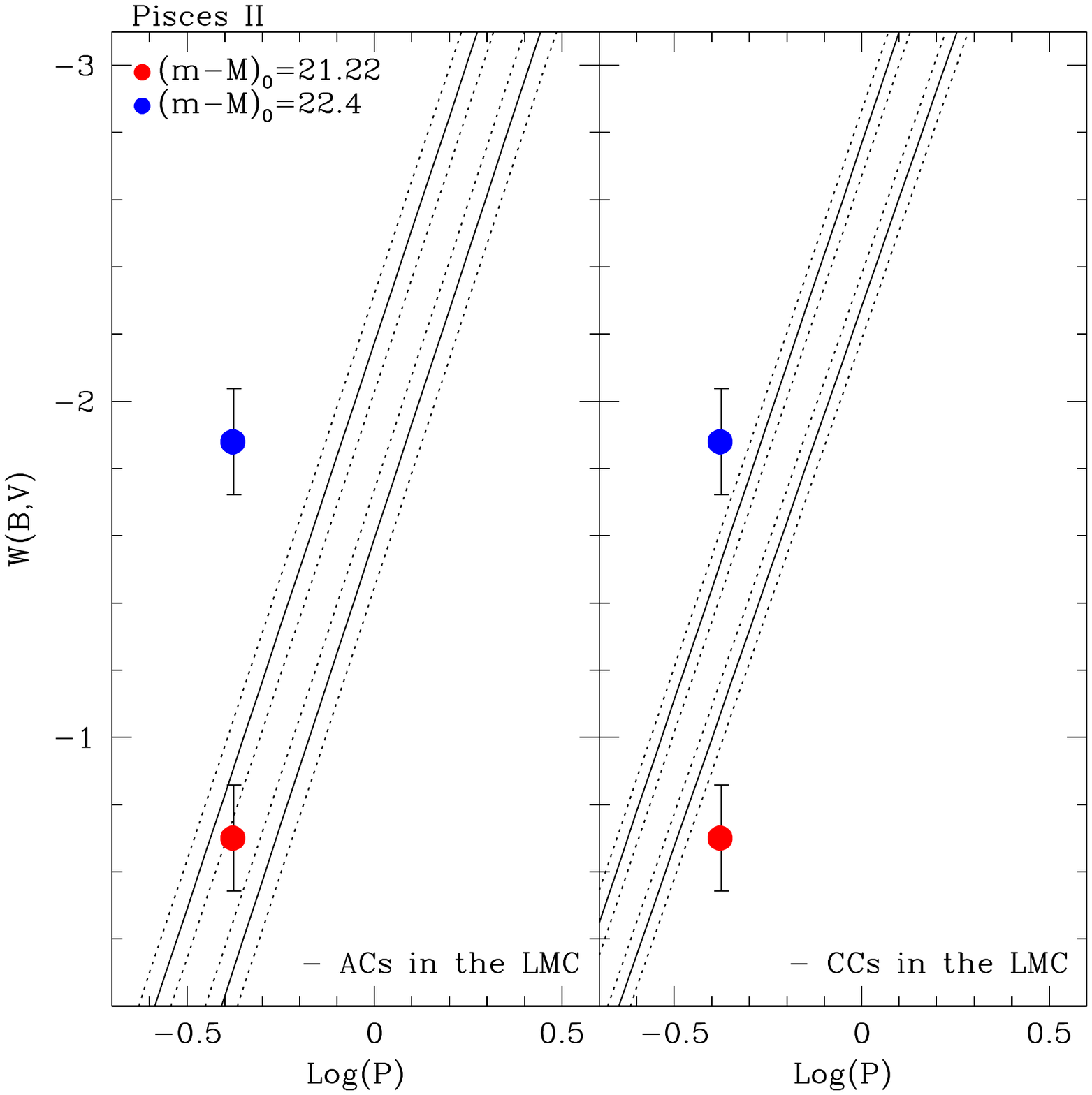}
\caption{$PW$ relations (solid lines) for ACs (left panel; from \citealt{rip14}) and CCs (right panel; from \citealt{jac16}) in the LMC, transformed to the $B$ and $V$ bands.  Dotted lines represent the 1$\sigma$ uncertainties, which for the AC relations correspond to 0.15 mag and for the CCs to 0.10 mag. A red filled circle shows the position of V4, according to the period of $\sim$ 0.42 d and a Wesenheit index calculated for a distance modulus of $(m-M)_{0}$ = 21.22 mag as estimated from V1, the RRL inside the half-light radius of Psc~II.  The blue filled circle show V4 but adopting a mass of 1.8M$_{\bigodot}$ and a Wesenheit index calculated for a distance modulus of 22.4 mag.}
\label{fig:pw_psc2_a}
\end{figure}

\begin{figure}[hbp]
\centering
\includegraphics[trim= 60 160 10 80 clip, width=9cm]{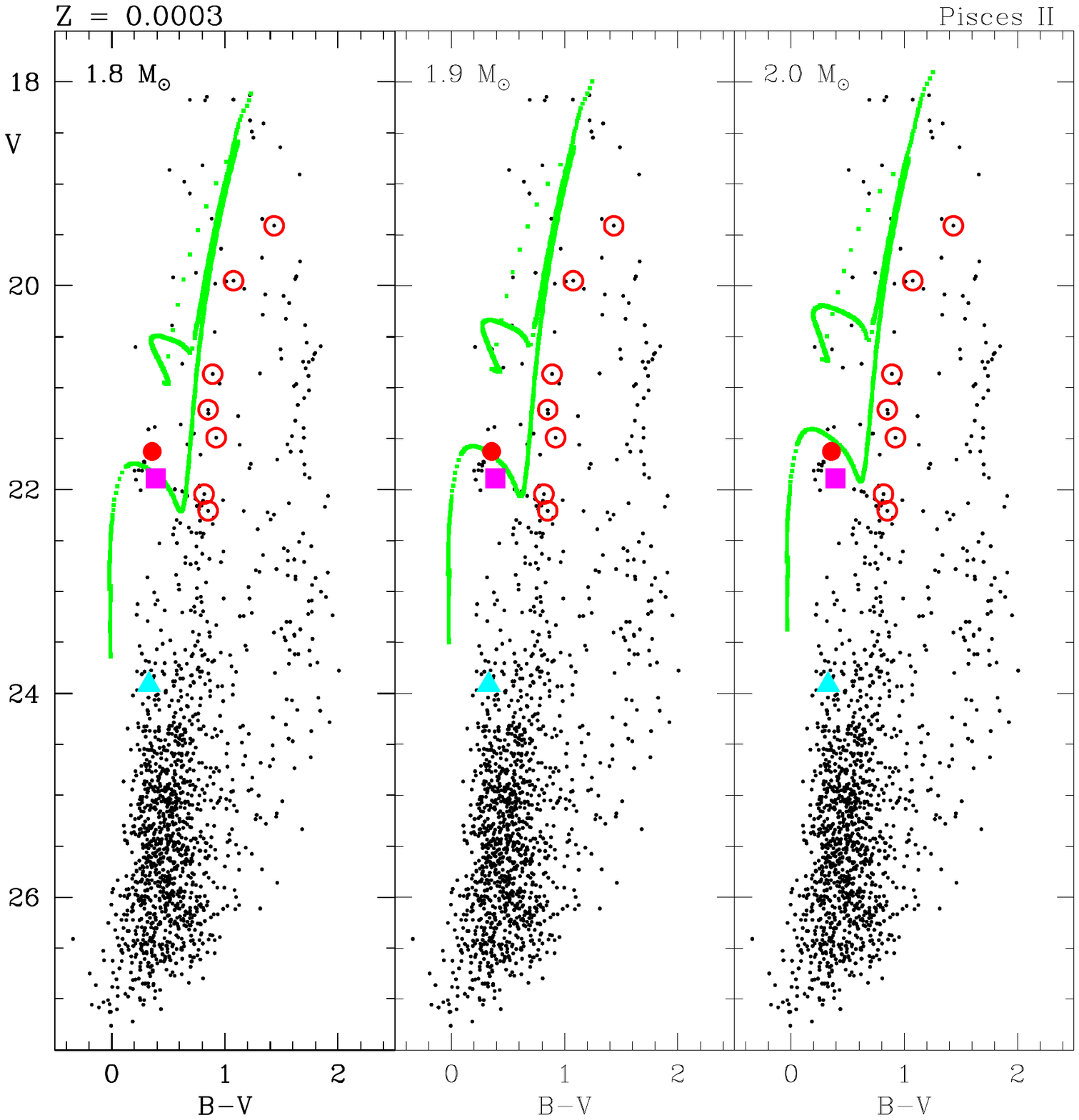}
\caption{Stellar evolutionary tracks (BaSTI web interface) for 1.8, 1.9 and 2.0 M$_{\bigodot}$ (green solid line; left, central and right panels, respectively) and metallicity  Z = 0.0003 ([Fe/H] = $-$1.8 dex) overlaid on the CMD of Psc~II. The tracks were corrected adopting the distance modulus of Psc~II derived from the RRab star [V1; $(m-M)_{0}$ =21.22 mag] and the reddening E($B-V$)=0.056 from \citet{sef11} maps. V1 is marked by a magenta filled square, the AC/RRab star (V4) by a filled red circle, the SX Phe star (V3) by a cyan triangle and the spectroscopically confirmed members from \citet{kir15} by red open circles.}
\label{fig:track_psc2}
\end{figure}

\begin{figure}[hbp]
\centering
\includegraphics[trim= 20 150 0 100 clip, width=10cm]{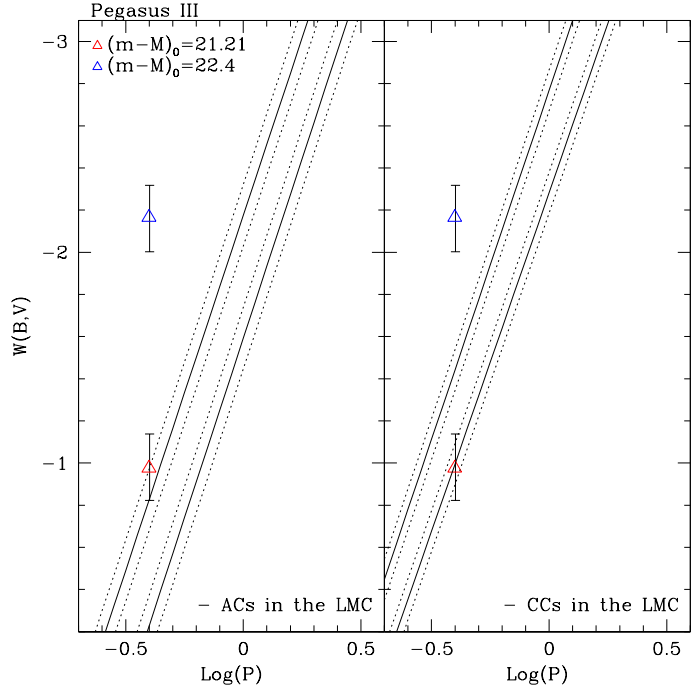}
\caption{Same as in Fig.~\ref{fig:pw_psc2_a} but for the variable star with uncertain type classification in Peg~III, star V2. A red triangle shows the position of V2, according to the period as AC (0.40 d) and a Wesenheit index calculated for a distance modulus of $(m-M)_{0}$ = 21.21 mag, as inferred from V1, the bona fide RRab star in Peg~III. The blue triangle shows V2 but adopting a mass of 1.8 M$_{\bigodot}$ and a Wesenheit index calculated for a distance modulus of 22.4 mag.}
\label{fig:pw_peg3_a}
\end{figure}

\begin{figure}[hbp]
\centering
\includegraphics[trim= 20 130 0 120 clip, width=9cm]{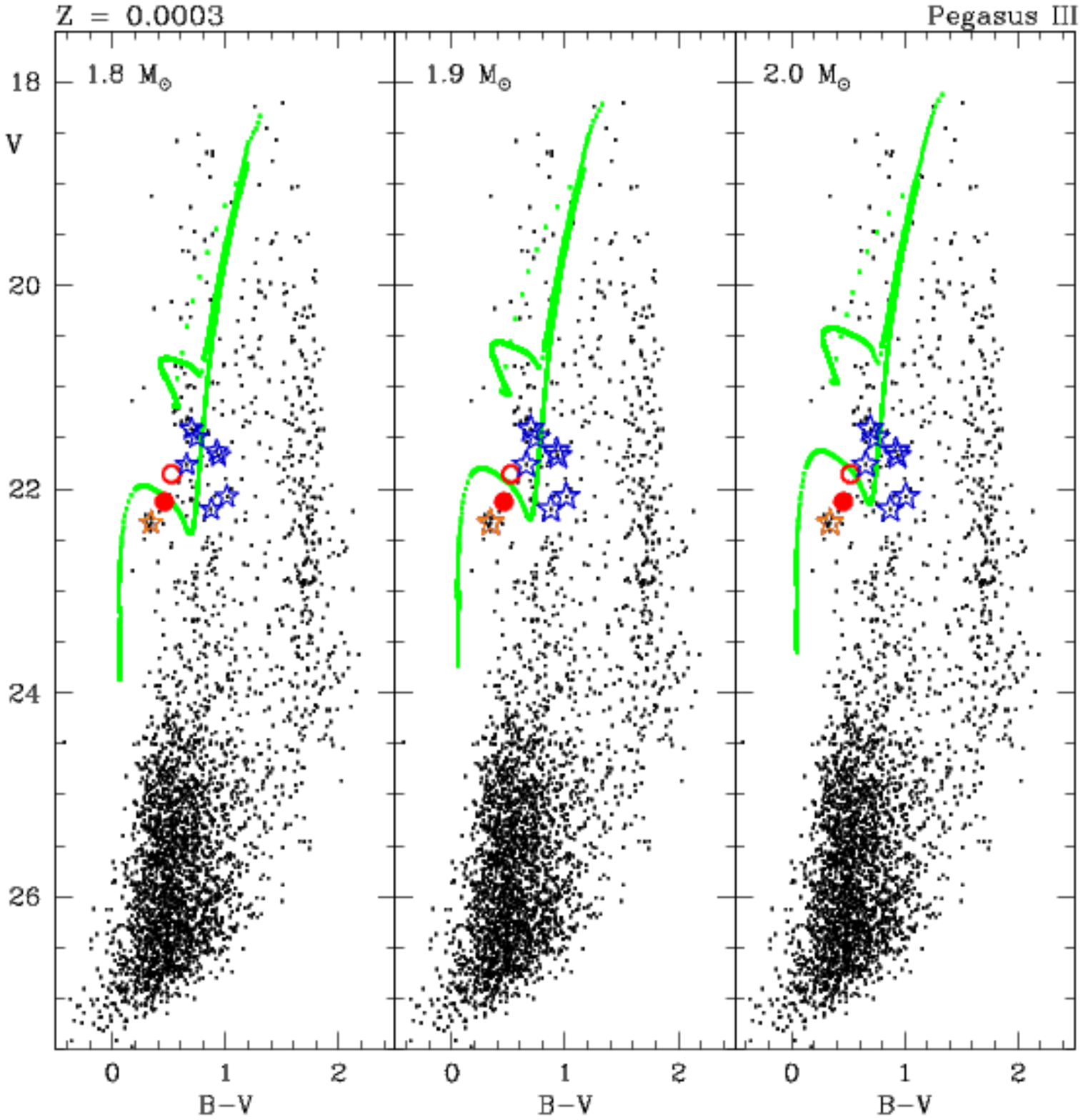}
\caption{Stellar evolutionary tracks (BaSTI web interface) for 1.8, 1.9 and 2.0 M$_{\bigodot}$ (green solid line; left, central and right panels, respectively) and metallicity  Z = 0.0003 ([Fe/H] = $-$1.8 dex) overlaid on the CMD of Peg~III. The tracks were corrected adopting the distance modulus of Peg~III derived from the RRab star [V1; $(m-M)_{0}$ =21.21 mag] and the reddening E($B-V$)=0.13 from \citet{sef11} maps. V2 is marked by a red open circle, the bona-fide RRab star V1 by a red filled circle and the spectroscopically confirmed members from \citet{kim16} by blue and orange stars.}
\label{fig:track_peg3}
\end{figure}


In Peg~III we have identified 2 variable stars (lower portion of Table~\ref{tab:psc2_var}): an RRab star with P $\sim$ 0.55 days (V1) and a variable with uncertain type classification (V2). Both variables are located outside the galaxy  r$_{h}$, 
respectively within 4r$_{h}$ and 10r$_{h}$ from the center of Peg~III. Similarly to  Psc~II the period and amplitude of the  bona-fide RRab star (V1) 
place the variable star 
close to the Oo~I locus in the period–amplitude diagram  (Figure~\ref{fig:bai_psc2_1}). 
According to the distance modulus 
inferred from  V1,  21.21 $\pm$ 0.23 mag, Peg~III is at 174  $\pm$ 18 kpc, in good agreement with the literature and at the same distance from us as Psc~II. 
V2 in Peg~III has totally similar characteristic  as V4 in Psc~II. It could either be an RRab star with P $\sim$ 0.66 d or an AC with P $\sim$ 0.40 d and about  0.3 mag brighter than the galaxy  HB. However,  similar tests as done for Psc~II-V4, lead us to  
conclude that V2 is very likely an RRab star with Oo~II characteristics belonging to Peg~III. 

The comparison of the 
CMD with stellar isochrones  suggests 
the presence also in Peg~III of a dominant old stellar component ($>$ 10 Gyr) with metallicity larger than [Fe/H] $\sim -$1.8 dex. However, the RGB of Peg~III exhibits a non negligible spread in color,  
indicative of a significant metallicity spread. This evidence, along with 
the 
pulsation properties of V2, indicate the presence  in Peg~III of an old, more metal-poor ([Fe/H] $\sim-$ 2.4 dex)  stellar component, as also supported by the comparison with the mean ridge lines of Galactic GCs (see Fig.~\ref{fig:cmd_psc2_rl}). 

 In summary, Psc~II and Peg~III not only share very similar properties but also almost same issues. 
There are a number of alternative hypotheses that could explain why Psc~II and Peg~III have so much similar properties:
1) they could be the leftovers of a same large galaxy, which entered the MW environment and was totally disrupted by a close encounter with our Galaxy some Gyrs ago;
2) they could be satellites of one of the present day MW dSph satellites or of a dSph that has been completely cannibalized by the MW;
3) they could be separate entities, that just by chance are close on the sky and share similar properties;
and, finally:
4) they could have been born in a pair.
However, in spite of all the similarities and the proximity on the sky, the density contour maps do not reveal signatures of a tidal interaction between them. 
If they were indeed born in a pair, or were  physically connected in the past, no trace of this physical connection seems to have survived.
\par 
The analysis we have presented here revealed very intriguing characteristics and similarities of Psc~II and Peg~III, prompting  us to further study these systems. 
Unfortunately, the HB of the two  galaxies is at/below the limiting magnitude reached by \textit{Gaia} ($V\sim$ 20.5-21 mag; \citealt{gai18b}); but their  brighter RGB stars 
are well within the reach of \textit{Gaia}.
Psc~II and Peg~III are in a region of the sky still poorly sampled by the  observations released in \textit{Gaia} DR2. However, the \textit{Gaia} sky coverage is rapidly increasing  and in future data releases \textit{Gaia} will likely tell us whether Psc~II and Peg~III  share the same relative proper motions.
Furthermore, in the near future the Rubin Observatory Large Synoptic Survey Telescope (LSST), designed to collect deep (5 mag deeper than \textit{Gaia}) ground-based, wide-field imaging from the blue to the near-infrared wavelengths, will be able to provide observations of Psc~II and Peg~III and the field between them deep enough to shed light on the nature and origin of these intriguing couple.\\

\appendix

\section{Psc~II-V4 and  Peg~III-V2 as Anomalous Cepheids}\label{sec:appendixA}

ACs are known to follow specific Period-Lumiosity ($PL$) and Period-Wesenheit ($PW$) relations, which differ from both the classical and the Type II Cepheids relationships \citep{sos08}. Therefore, if V4 and V2 are ACs, they should conform to the $PL$ and $PW$ relations of ACs.
\par
We have compared V4 with the $PW$ relations 
followed by ACs in the Large Magellanic Cloud \citep[LMC; see][]{rip14}. These relations are defined in the $V$ and $I$ bands, hence we converted them to the $B$, $V$ bands\footnote{The Wesenheit index in our case is $W(B, V)= M_{V} - 3.1\times(B-V)$, where M$_{V}$ is the $V$ magnitude corrected for the distance.} using the relation $\langle W(V,I) \rangle =-0.22 (\pm 0.001) + 1.03 \langle W(B,V) \rangle$ (Eq.~12 of \citealt{marc04}).

 The $PW$ relations of the LMC ACs converted into $B$ and $V$ bands are shown as solid lines in the left panel of the Fig.~\ref{fig:pw_psc2_a}, together with their 1$\sigma$ uncertainties (dotted lines). In this figure the lower line is the $PW$ relation for fundamental mode ACs, while the upper line is for ACs that pulsate in the first-overtone mode.
Psc~II-V4 (red filled circle) is plotted according to the  period as an AC and the Wesenheit index computed by correcting the  star $\langle V \rangle $-band apparent magnitude for the distance modulus (m-M)$_0$ = 21.22 mag as estimated from V1, the RRab star within the r$_{h}$ of Psc~II. V4 lies within the 1$\sigma$ uncertainty of the $PW$ relation for first-overtone ACs, hence, it could indeed be an AC belonging to Psc~II. As a check, in the right panel of  Fig.~\ref{fig:pw_psc2_a}  we have compared V4 also with the $PW$ relations for LMC Classical Cepheids (CCs) by \citet{jac16}. The star appears to be significantly fainter than 
the Wesenheit index of a fundamental-mode CCs with same period, hence,  
it is  unlikely that it could be a CC of Psc~II. Furthermore, in the CMD of Psc~II (see Fig.~\ref{fig:cmd_psc2_5r}) 
there is no indication of the presence of a young population ($<$ 500 Myr) as that traced by CCs [for a detailed description of the Psc~II CMD (and the Peg~III CMD as well) we refer the reader to Section~\ref{sec:psc2_cmd}]. 
Figure~\ref{fig:pw_psc2_a} seems thus to support the possibility that V4 might be a first overtone AC of Psc~II. We further checked this hypothesis comparing the star with stellar evolutionary tracks by \citet[available at \url{http://basti.oa- teramo.inaf.it/index.html}]{pie04}, for the typical mass and metallicity of ACs ($\sim$1.8-2.2 M$_{\bigodot}$ according to \citealt{cap04} and Z = 0.0004 following \citealt{marc04}).

This is illustrated in the three panels of Fig.~\ref{fig:track_psc2} that show the CMD of Psc~II with overlaid evolutionary tracks for metallicity Z = 0.0003 and masses of 1.8, 1.9 and 2.0 M$_{\bigodot}$, corrected for the distance modulus of Psc~II and for a reddening of E($B-V$) = 0.056 mag according to \citet{sef11}. 

 This comparison shows that if V4 were  indeed an AC belonging to Psc~II it should be at least 1 mag brighter to lie on the proper region of the evolutionary track: the hook above 
the HB. On the other hand, since V4 is at about five r$_{h}$ from the Psc~II center, it could  be a field AC. We thus shifted the stellar tracks as to fit the luminosity of V4 and  found that the hook of the 1.8 M$_{\bigodot}$ track would fit the luminosity of V4 for a distance modulus of $(m-M)_{0}$ $\sim$ 22.4 mag and for longer moduli in case of larger masses.

However, adopting such longer distance moduli, V4 could no longer fit any  of the $PW$ relations for ACs or CCs, as clearly shown by blue filled circles in Fig.~\ref{fig:pw_psc2_a}. Based on these conflicting results we conclude that the classification of V4 as an AC either belonging to Psc~II or to the background field is very unlikely\footnote{ We have shown in Sect.~\ref{sec:v4_psc2} that the probability of finding  field RRLs at the distance of Psc~II and Peg~III is very much low. Finding MW field ACs, at such distances is even less probable,  because ACs are much rarer than RRLs. Indeed, within a region of 20,000 deg$^2$ covered by the Catalina Surveys Data Release-1, \citet{dra14} have identified a total number of only 64 candidate ACs and a small fraction of them, no more than $\sim0.01-0.015$, is distributed beyond Psc~II and Peg~III Galactic latitudes (l $>$ 65$^\circ$). Since the LBT FoV is more than 10$^6$ times smaller than the area covered by \citet{dra14}, the fraction of MW halo ACs at the Galactic latitude of Psc~II and Peg~III is approximately $9\times10^{-8}$.}. Totally similar conflicting results were found also for V2 in Peg~III  (see Figs.~\ref{fig:pw_peg3_a} and~\ref{fig:track_peg3}), 
leading us to rule out a classification as AC also for V2 in Peg~III.

\section{DOTTER et AL. 2008 ISOCHRONES ON THE CMD of Psc~II}\label{sec:appendixB}

\begin{figure}[hbp]
\centering
\includegraphics[trim= 20 110 0 100 clip, width=12cm]{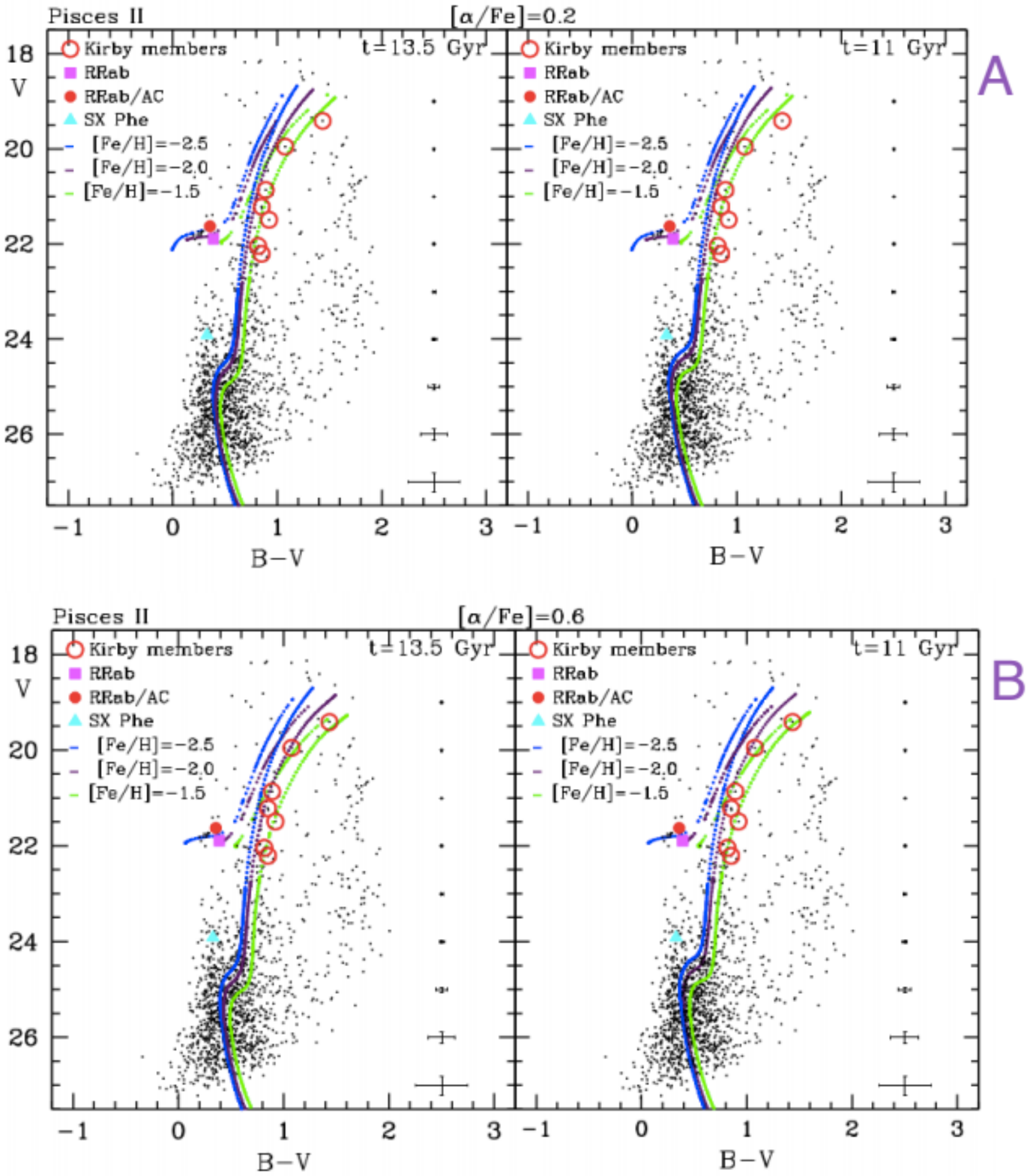}
\caption{Same as left panel of Fig~\ref{fig:cmd_psc2_5r}  with overlaid  \citet{dot08}'s isochrones for different metallicities (from $-$2.5 to $-$1.5 dex), a fixed age (13.5 Gyr left panels; 11 Gyr right panels) and a fixed [$\alpha$/Fe] enhancement of 0.2 dex (fig. A) and 0.6 dex (fig. B).}
\label{fig:dot}
\end{figure}

 Figure~\ref{fig:dot} shows CMD of Psc~II (same as left panel of Fig~\ref{fig:cmd_psc2_5r}) with overlaid  \citet{dot08}'s isochrones for different metallicities, $-$2.5, $-$2.0 and $-$1.5 dex, a fixed age (13.5 Gyr left panels; 11Gyr right panels) and a fixed [$\alpha$/Fe] enhancement of 0.2 dex in fig A  and
0.6 dex in fig. B.
The best agreement between variable stars and spectroscopic metal-poor stars is achieved for an isochrone with metal abundance between $-$2  and $-$1.5 dex and rather old.  Here,  \citet{dot08} isochrones with 13.5 Gyr seem to be quite consistent with the PARSEC isochrones of 13 Gyr we have used throughout the paper. In summary, results do not seem to change significantly changing the isochrone set adopted as reference. On the other hand, the discrepancy with previous studies may also arise because the photometry in this paper is much deeper than previous works.

\acknowledgments

 The authors thank the anonymous referee for her/his detailed comments, which have helped us to make the paper much clearer and readable.
We warmly thank P. Montegriffo for the development and maintenance of the GRATIS software. AG thanks M. Catelan for useful comments on an earlier version of this work. Financial support for this research was provided by
the Italian Ministry of University and Research through
 the Premiale project (2015) ``MITiC: MIning The Cosmos Big Data and Innovative Italian Technology for Frontier Astrophysics and Cosmology". The LBT is an international collaboration among institutions in the United States, Italy, and Germany. LBT Corporation partners are The University of Arizona on behalf of the Arizona university system; Istituto Nazionale di Astrofisica, Italy; LBT Beteiligungsgesellschaft, Germany, representing the Max-Planck Society, the Astrophysical Institute Potsdam, and Heidelberg University; The Ohio State University; and The Research Corporation, on behalf of The University of Notre Dame, University of Minnesota, and University of Virginia. We acknowledge the support from the LBT-Italian Coordination Facility for the execution of observations, data distribution, and reduction.

\end{document}